\documentclass[twocolumn,twocolappendix,resetfootnote]{aastex701}

\usepackage[varvw,scale=1.08]{newtx}
\usepackage[T1]{fontenc}

\usepackage{lipsum}
\usepackage{enumitem}
\usepackage{physics}
\usepackage{multirow}
\usepackage{textgreek}
\usepackage{bm}

\hypersetup{pdfnewwindow=true,
     colorlinks=true,
     linkcolor=blue,
     citecolor=blue,
	 final=true}
	 
\usepackage{etoolbox}
\makeatletter
\patchcmd{\NAT@citex}
  {\@citea\NAT@hyper@{%
     \NAT@nmfmt{\NAT@nm}%
     \hyper@natlinkbreak{\NAT@aysep\NAT@spacechar}{\@citeb\@extra@b@citeb}%
     \NAT@date}}
  {\@citea\NAT@nmfmt{\NAT@nm}%
   \NAT@aysep\NAT@spacechar\NAT@hyper@{\NAT@date}}{}{}
\patchcmd{\NAT@citex}
  {\@citea\NAT@hyper@{%
     \NAT@nmfmt{\NAT@nm}%
     \hyper@natlinkbreak{\NAT@spacechar\NAT@@open\if*#1*\else#1\NAT@spacechar\fi}%
       {\@citeb\@extra@b@citeb}%
     \NAT@date}}
  {\@citea\NAT@nmfmt{\NAT@nm}%
   \NAT@spacechar\NAT@@open\if*#1*\else#1\NAT@spacechar\fi\NAT@hyper@{\NAT@date}}
  {}{}
\makeatother

\newcommand{\JWST}{\textit{JWST}}

\newcommand{\target}{CW-LRD-z10}

\begin{document}

\title{\large EMBER: A Little Red Dot Candidate at $\bm{z_{\rm spec}=10.36}$ in the Seed Black Hole Era
}
\shortauthors{Tanaka, Akins, et al.}
\shorttitle{EMBER: CW-LRD-z10 at $z_{\rm spec}=10.36$}

\suppressAffiliations

\author[0009-0003-4742-7060]{Takumi S. Tanaka}
\affiliation{Kavli Institute for the Physics and Mathematics of the Universe (WPI), The University of Tokyo Institutes for Advanced Study, The University of Tokyo, Kashiwa, Chiba 277-8583, Japan}
\affiliation{Department of Astronomy, Graduate School of Science, The University of Tokyo, 7-3-1 Hongo, Bunkyo-ku, Tokyo 113-0033, Japan}
\affiliation{Center for Data-Driven Discovery, Kavli IPMU (WPI), UTIAS, The University of Tokyo, Kashiwa, Chiba 277-8583, Japan}
\email{takumi.tanaka@ipmu.jp}

\author[0000-0003-3596-8794]{Hollis B. Akins}
\altaffiliation{Einstein Fellow}
\affiliation{Department of Astrophysical Sciences, Princeton University, 4 Ivy Lane, Princeton, NJ 08544, USA}
\email{hollis.akins@gmail.com}

\author[0000-0002-0000-6977]{John D. Silverman}
\affiliation{Kavli Institute for the Physics and Mathematics of the Universe (WPI), The University of Tokyo Institutes for Advanced Study, The University of Tokyo, Kashiwa, Chiba 277-8583, Japan}
\affiliation{Department of Astronomy, Graduate School of Science, The University of Tokyo, 7-3-1 Hongo, Bunkyo-ku, Tokyo 113-0033, Japan}
\affiliation{Center for Data-Driven Discovery, Kavli IPMU (WPI), UTIAS, The University of Tokyo, Kashiwa, Chiba 277-8583, Japan}
\affiliation{Center for Astrophysical Sciences, Department of Physics \& Astronomy, Johns Hopkins University, Baltimore, MD 21218, USA}
\email{john.silverman@ipmu.jp}

\author[0000-0002-0930-6466]{Caitlin M. Casey}
\affiliation{Department of Physics, University of California, Santa Barbara, CA 93106, USA}
\affiliation{Cosmic Dawn Center (DAWN), Denmark}
\email{cmcasey@ucsb.edu}

\author[0000-0001-7568-6412]{Ezequiel Treister}
\affiliation{Instituto de Alta Investigaci\'on, Universidad de Tarapac\'a, Casilla 7D, Arica, Chile}
\email{etreister@academicos.uta.cl}

\author[0000-0002-6290-3198]{Manuel Aravena}
\affiliation{Instituto de Estudios Astrof\'{\i}cos, Facultad de Ingenier\'{\i}a y Ciencias, Universidad Diego Portales, Av. Ej\'ercito 441, Santiago, Chile}
\affiliation{Millenium Nucleus for Galaxies (MINGAL)}
\email{manuel.aravenaa@mail.udp.cl}

\author[0000-0002-6047-430X]{Yuichi Harikane}
\affiliation{Institute for Cosmic Ray Research, The University of Tokyo, 5-1-5 Kashiwanoha, Kashiwa, Chiba 277-8582, Japan}
\email{hari@icrr.u-tokyo.ac.jp}

\author[0000-0002-3683-7297]{Benny Trakhtenbrot}
\affiliation{School of Physics and Astronomy, Tel Aviv University, Tel Aviv 69978, Israel}
\email{benny.trakht@gmail.com}

\author[0000-0001-9840-4959]{Kohei Inayoshi}
\affiliation{Kavli Institute for Astronomy and Astrophysics, Peking University, Beijing 100871, China}
\email{inayoshi.pku@gmail.com}

\author[0000-0002-2597-2231]{Kazuhiro Shimasaku}
\affiliation{Department of Astronomy, Graduate School of Science, The University of Tokyo, 7-3-1 Hongo, Bunkyo-ku, Tokyo 113-0033, Japan}
\affiliation{Research Center for the Early Universe, Graduate School of Science, The University of Tokyo, 7-3-1 Hongo, Bunkyo-ku,
Tokyo 113-0033, Japan}
\email{shimasaku@astron.s.u-tokyo.ac.jp}

\author[0000-0002-3560-8599]{Maximilien Franco}
\affiliation{Universit\'e Paris-Saclay, Universit\'e Paris Cit\'e, CEA, CNRS, AIM, 91191, Gif-sur-Yvette, France}
\email{francomaximilien.astro@gmail.com}

\author[0000-0001-7201-5066]{Seiji Fujimoto}
\affiliation{David A. Dunlap Department of Astronomy and Astrophysics, University of Toronto, 50 St. George Street, Toronto, Ontario, M5S 3H4, Canada}
\affiliation{Dunlap Institute for Astronomy and Astrophysics, 50 St. George Street, Toronto, Ontario, M5S 3H4, Canada}
\email{seiji.fujimoto@utoronto.ca}

\author[0000-0001-6251-649X]{Mingyu Li}
\affiliation{Department of Astronomy, Tsinghua University, Beijing 100084, China}
\affiliation{Kavli Institute for Cosmology, University of Cambridge, Madingley Road, Cambridge CB3 0HA, UK}
\affiliation{Cavendish Laboratory, University of Cambridge, 19 JJ Thomson Avenue, Cambridge CB3 0HE, UK}
\email{lmy22@mails.tsinghua.edu.cn}

\author[0009-0008-0167-5129]{Hiroya Umeda}
\affiliation{Center for Computational Sciences, University of Tsukuba, Ten-nodai, 1-1-1 Tsukuba, Ibaraki 305-8577, Japan}
\email{ume@icrr.u-tokyo.ac.jp}

\author[0000-0002-4544-8242]{Jan-Torge Schindler}
\affiliation{Hamburger Sternwarte, University of Hamburg, Gojenbergsweg 112, D-21029 Hamburg, Germany}
\email{jan-torge.schindler@uni-hamburg.de}

\author[0000-0002-7779-8677]{Akio K.\ Inoue}
\affiliation{Department of Pure and Applied Physics, Graduate School of Advanced Science and Engineering, Faculty of Science and Engineering, Waseda University, 3-4-1 Okubo, Shinjuku, Tokyo 169-8555, Japan}
\affiliation{Waseda Research Institute for Science and Engineering, Faculty of Science and Engineering, Waseda University, 3-4-1 Okubo, Shinjuku, Tokyo 169-8555, Japan}
\email{akinoue@aoni.waseda.jp}

\author[0009-0004-4168-3634]{Jean-Baptiste Billand}
\affiliation{Universit\'e Paris-Saclay, Universit\'e Paris Cit\'e, CEA, CNRS, AIM, 91191, Gif-sur-Yvette, France}
\email{Jean-Baptiste.BILLAND@cea.fr}

\author[0000-0002-2603-2639]{Darshan Kakkad}
\affiliation{Centre for Astrophysics Research, University of Hertfordshire, Hatfield, AL10 9AB, UK}
\email{darshankakkad@gmail.com}

\author[0000-0001-8917-2148]{Xuheng Ding}
\affiliation{School of Physics and Technology, Wuhan University, Wuhan 430072, China}
\email{dingxh@whu.edu.cn}

\author[0000-0001-8519-1130]{Steven L. Finkelstein}
\affiliation{Department of Astronomy, The University of Texas at Austin, Austin, TX, USA}
\affiliation{Cosmic Frontier Center, The University of Texas at Austin, Austin, TX, USA}
\email{stevenf@astro.as.utexas.edu}

\author[0000-0002-3301-3321]{Michaela Hirschmann}
\affiliation{Institute of Physics, GalSpec, Ecole Polytechnique Federale de Lausanne, Observatoire de Sauverny, Chemin Pegasi 51, 1290 Versoix, Switzerland}
\affiliation{INAF, Astronomical Observatory of Trieste, Via Tiepolo 11, 34131 Trieste, Italy}
\email{michaela.hirschmann@epfl.ch}

\author[0000-0002-6467-8078]{Z.~Ghaffari}
\affiliation{INAF Osservatorio Astronomico di Trieste, Via G.B. Tiepolo 11 e via Bazzoni 2. 34100, Trieste, Italy \label{inaft}}
\affiliation{IFPU: Institute for Fundamental Physics of the Universe, Via Beirut, 2, 34151, Trieste, Italy \label{ifpu}}
\email{zohreh.ghaffari@inaf.it}

\author[0000-0003-4569-2285]{Andrew J. Battisti}
\affiliation{International Centre for Radio Astronomy Research, University of Western Australia, 7 Fairway, Crawley, WA 6009, Australia}
\affiliation{Research School of Astronomy and Astrophysics, Australian National University, Cotter Road, Weston Creek, ACT 2611, Australia}
\email{Andrew.Battisti@anu.edu.au}

\author[0000-0002-3473-6716]{Lauro Moscardini}
\affiliation{University of Bologna - Department of Physics and Astronomy “Augusto Righi” (DIFA), Via Gobetti 93/2, I-40129 Bologna, Italy}
\affiliation{INAF- Osservatorio di Astrofisica e Scienza dello Spazio, Via Gobetti 93/3, I-40129, Bologna, Italy}
\affiliation{INFN- Sezione di Bologna, Viale Berti Pichat 6/2, I-40127 Bologna, Italy}
\email{lauro.moscardini@unibo.it}

\author[0000-0003-4761-2197]{Nicole E. Drakos}
\affiliation{Department of Physics and Astronomy, University of Hawaii, Hilo, 200 W Kawili St, Hilo, HI 96720, USA}
\email{ndrakos@hawaii.edu}

\author[0000-0003-2716-8332]{Rasha M. Samir}
\affiliation{Department of Astronomy, National Research Institute of Astronomy and Geophysics (NRIAG), Cairo, 11421, Egypt} 
\email {rasha.samir@nriag.sci.eg}

\author[0000-0001-9187-3605]{Jeyhan S. Kartaltepe}
\affiliation{Laboratory for Multiwavelength Astrophysics, School of Physics and Astronomy, Rochester Institute of Technology, 84 Lomb Memorial Drive, Rochester, NY 14623, USA}
\email{jeyhan@astro.rit.edu}

\author[0000-0001-7232-5355]{Qinyue Fei}
\affiliation{David A. Dunlap Department of Astronomy and Astrophysics, University of Toronto, 50 St. George Street, Toronto, Ontario, M5S 3H4, Canada}
\email{qy.fei@utoronto.ca}

\author[0000-0001-5105-2837]{Ming-Yang Zhuang}
\affiliation{Department of Astronomy, University of Illinois Urbana-Champaign, Urbana, IL 61801, USA}
\email{mingyang@illinois.edu}

\author[0000-0002-7303-4397]{Olivier Ilbert}
\affiliation{Aix Marseille Univ, CNRS, CNES, LAM, Marseille, France}
\email{olivier.ilbert@lam.fr}

\author[0000-0002-7964-6749]{Wuji Wang}
\affiliation{IPAC, California Institute of Technology, 1200 E. California Blvd. Pasadena, CA 91125, USA}
\email{wujiwang@ipac.caltech.edu}

\author[0000-0002-0569-5222]{Rafael C. Arango-Toro}
\affiliation{Institut d'Astrophysique de Paris, UMR 7095, CNRS, and Sorbonne Université, 98bis boulevard Arago, 75014 Paris, France}
\email{arango@iap.fr}

\author[0000-0002-4205-9567]{Hiddo S. B. Algera} 
\affiliation{Institute of Astronomy and Astrophysics, Academia Sinica, 11F of Astronomy-Mathematics Building, No.1, Section 4, Roosevelt Rd, Taipei 106319, Taiwan, R.O.C}
\email{hsbalgera@asiaa.sinica.edu.tw}

\author[0009-0005-3133-1157]{Greta Toni}
\affiliation{University of Bologna - Department of Physics and Astronomy “Augusto Righi” (DIFA), Via Gobetti 93/2, I-40129 Bologna, Italy}
\affiliation{INAF- Osservatorio di Astrofisica e Scienza dello Spazio, Via Gobetti 93/3, I-40129, Bologna, Italy}
\affiliation{Zentrum f\"{u}r Astronomie, Universit\"{a}t Heidelberg, Philosophenweg 12, D-69120, Heidelberg, Germany}
\email{greta.toni4@unibo.it}

\author[0009-0007-0864-7094]{Junya Arita}
\affiliation{Department of Astronomy, Graduate School of Science, The University of Tokyo, 7-3-1 Hongo, Bunkyo-ku, Tokyo 113-0033, Japan}
\email{jarita@astron.s.u-tokyo.ac.jp}

\author[0000-0002-9453-0381]{Kei Ito}
\affiliation{Cosmic Dawn Center (DAWN), Denmark}
\affiliation{DTU Space, Technical University of Denmark, Elektrovej 327 DK2800 Kgs. Lyngby, Denmark}
\email{keiitoastro@gmail.com}

\author[0000-0002-0236-919X]{Ghassem Gozaliasl}
\affiliation{Department of Computer Science, Aalto University, P.O. Box 15400, FI-00076 Espoo, Finland}
\affiliation{Department of Physics, University of Helsinki, P.O. Box 64, FI-00014 Helsinki, Finland}
\email{ghassemgozaliasl@gmail.com}

\author[0000-0002-8360-3880]{Dale D. Kocevski}
\affiliation{Department of Physics and Astronomy, Colby College, Waterville, ME 04901, USA}
\email{dkocevsk@colby.edu}

\author[0000-0003-0111-8249]{Yunjing Wu}
\affiliation{Kavli Institute for the Physics and Mathematics of the Universe (WPI), The University of Tokyo Institutes for Advanced Study, The University of Tokyo, Kashiwa, Chiba 277-8583, Japan}
\affiliation{Center for Data-Driven Discovery, Kavli IPMU (WPI), UTIAS, The University of Tokyo, Kashiwa, Chiba 277-8583, Japan}
\email{yunjing.wu@ipmu.jp}

\author[0000-0003-2984-6803]{Masafusa Onoue}
\affiliation{Waseda Institute for Advanced Study (WIAS), Waseda University, 1-21-1, Nishi-Waseda, Shinjuku, Tokyo 169-0051, Japan}
\affiliation{Kavli Institute for the Physics and Mathematics of the Universe (WPI), The University of Tokyo Institutes for Advanced Study, The University of Tokyo, Kashiwa, Chiba 277-8583, Japan}
\email{masafusa.onoue@aoni.waseda.jp}

\author[0000-0002-1416-8483]{Marc Huertas-Company}
\affiliation{Instituto de Astrofísica de Canarias (IAC), La Laguna, E-38205, Spain }
\affiliation{Universidad de La Laguna. Avda. Astrofísico Fco. Sanchez, La Laguna, Tenerife, Spain}
\affiliation{Observatoire de Paris, LUX, PSL University, 61 avenue de l’Observatoire, F-75014 Paris, France}
\affiliation{Université de Paris, 5 Rue Thomas Mann, 75014 Paris, France}
\email{mhuertas@iac.es}

\author[0009-0006-8707-023X]{Yiyang Zhang}
\affiliation{School of Physics and Technology, Wuhan University, Wuhan 430072, China}
\email{yiyang_zhang@whu.edu.cn}

\author[0000-0002-4485-8549]{Jason Rhodes}
\affiliation{Jet Propulsion Laboratory, California Institute of Technology, 4800 Oak Grove Drive, Pasadena, CA 91001, USA}
\email{jason.d.rhodes@jpl.nasa.gov}

\author[0000-0002-9382-9832]{Andreas L. Faisst}
\affiliation{IPAC, California Institute of Technology, 1200 E. California Blvd. Pasadena, CA 91125, USA}
\email{afaisst@caltech.edu}

\author[0000-0001-5951-459X]{Zihao Li}
\affiliation{Cosmic Dawn Center (DAWN), Denmark}
\affiliation{Niels Bohr Institute, University of Copenhagen, Jagtvej 128, DK2200 Copenhagen N, Denmark}
\email{zihao.li@nbi.ku.dk}

\author[0000-0001-9189-7818]{Crystal L. Martin}
\affiliation{Department of Physics, University of California, Santa Barbara, CA 93106, USA}
\email{cmartin@physics.ucsb.edu}

\author[0000-0002-4225-4477]{Makoto Ando}
\affiliation{Institute for Cosmic Ray Research, The University of Tokyo, 5-1-5 Kashiwanoha, Kashiwa, Chiba 277-8582, Japan}
\email{makoto.ando.astro@gmail.com}

\author[0000-0002-5588-9156]{Vasily Kokorev}
\affiliation{Kapteyn Astronomical Institute, University of Groningen, PO Box 800, 9700 AV Groningen, The Netherlands}
\email{vasily.kokorev.astro@gmail.com}

\author[0009-0001-3422-3048]{Sophie L. Newman}
\affiliation{Institute of Cosmology and Gravitation, University of Portsmouth, Dennis Sciama Building, Burnaby Road, Portsmouth, PO1 3FX, UK}
\email{sophie.newman@port.ac.uk}

\author[0000-0002-8437-6659]{Can Xu}
\affiliation{School of Astronomy and Space Science, Nanjing University, Nanjing, Jiangsu 210093, China}
\affiliation{Key Laboratory of Modern Astronomy and Astrophysics, Nanjing University, Ministry of Education, Nanjing 210093, China}
\email{canxu@smail.nju.edu.cn}

\author[0009-0009-7403-8603]{Suin Matsui}
\affiliation{Department of Astronomy, Graduate School of Science, The University of Tokyo, 7-3-1 Hongo, Bunkyo-ku, Tokyo 113-0033, Japan}
\email{smatsui@astron.s.u-tokyo.ac.jp}

\author[0009-0001-9170-3363]{Jialai Wang}
\affiliation{Department of Astronomy, University of Science and Technology of China, Hefei 230026, China}
\affiliation{School of Astronomy and Space Science, University of Science and Technology of China, Hefei 230026, China}
\affiliation{Kavli Institute for the Physics and Mathematics of the Universe (WPI), The University of Tokyo Institutes for Advanced Study, The University of Tokyo, Kashiwa, Chiba 277-8583, Japan}
\affiliation{Center for Data-Driven Discovery, Kavli IPMU (WPI), UTIAS, The University of Tokyo, Kashiwa, Chiba 277-8583, Japan}
\email{jialaiwang@mail.ustc.edu.cn}

\author[0009-0000-0272-5468]{Carter Flayhart}
\affiliation{Laboratory for Multiwavelength Astrophysics, School of Physics and Astronomy, Rochester Institute of Technology, 84 Lomb Memorial Drive, Rochester, NY 14623, USA}
\email{cbf2388@rit.edu}

\author[0009-0007-3673-4523]{Hossein Hatamnia}
\affiliation{Department of Physics and Astronomy, University of California, Riverside, 900 University Avenue, Riverside, CA 92521, USA}
\affiliation{Observatories of the Carnegie Institution for Science, 813 Santa Barbara St., Pasadena, CA 91101, USA}
\email{hossein.hatamnia@email.ucr.edu}

\author[0000-0002-0101-336X]{Ali Ahmad Khostovan}
\affiliation{Department of Physics and Astronomy, University of Kentucky, 505 Rose Street, Lexington, KY 40506, USA}
\affiliation{Laboratory for Multiwavelength Astrophysics, School of Physics and Astronomy, Rochester Institute of Technology, 84 Lomb Memorial Drive, Rochester, NY 14623, USA}
\email{akhostov@gmail.com}

\author[0000-0002-0322-6131]{Ronaldo Laishram}
\affiliation{National Astronomical Observatory of Japan, 2-21-1 Osawa, Mitaka, Tokyo 181-8588, Japan}
\email{ronaldo.laishram@nao.ac.jp}

\author[0000-0002-9883-1413]{Felix Martinez III}
\affiliation{Laboratory for Multiwavelength Astrophysics, School of Physics and Astronomy, Rochester Institute of Technology, 84 Lomb Memorial Drive, Rochester, NY 14623, USA}
\email{fm5957@rit.edu}

\author[0000-0002-1233-9998]{David B. Sanders}
\affiliation{Institute for Astronomy, University of Hawaii, 2680 Woodlawn Drive, Honolulu, HI 96822, USA}
\email{sandersd@hawaii.edu}

\author[0000-0003-3631-7176]{Sune Toft}
\affiliation{Cosmic Dawn Center (DAWN), Denmark} 
\affiliation{Niels Bohr Institute, University of Copenhagen, Jagtvej 128, DK2200 Copenhagen N, Denmark}
\email{sune@nbi.ku.dk}

\author[0000-0001-6477-4011]{Francesco Valentino}
\affiliation{Cosmic Dawn Center (DAWN), Denmark} 
\affiliation{DTU Space, Technical University of Denmark, Elektrovej 327 DK2800 Kgs. Lyngby, Denmark}
\email{fmava@dtu.dk}

\author[0000-0002-7051-1100]{Jorge Zavala} 
\affiliation{Department of Astronomy, University of Massachusetts Amherst, 710 N Pleasant Street, Amherst, MA 01003, USA}
\email{jzavala@umass.edu}

\collaboration{all}{(Affiliations can be found after the references)}


\begin{abstract} 
Little red dots (LRDs) are a newly discovered population that may trace rapidly growing black holes (BHs) and provide a link between seed BHs and luminous active galactic nuclei (AGNs).  However, no LRD had previously been spectroscopically confirmed at $z>10$, leaving this earliest phase observationally unexplored. Using JWST/NIRSpec spectroscopy from the EMBER program, we measure a spectroscopic redshift of $z_{\rm spec}=10.361^{+0.003}_{-0.002}$ for \target, the highest-redshift LRD candidate known to date. The redshift is constrained by a prominent Ly$\alpha$ break and multiple tentative emission lines, including C\,{\sc iv}\,$\lambda\lambda1549,1551$ and [O\,{\sc iii}]\,$\lambda4363$ detected at $S/N=3.9$ and $3.6$, respectively. Its extreme NIRCam-MIRI color of $m_{\rm F444W}-m_{\rm F770W}=1.82\pm0.1$, characteristic V-shaped continuum with a turnover near the Balmer limit, and continuum slopes of $\beta_{\rm UV}=-2.26^{+0.11}_{-0.11}$ and $\beta_{\rm opt}=1.17^{+0.74}_{-0.70}$ all support its LRD nature. A possible iron complex at $\lambda_{\rm rest}\sim0.24\,\mathchar`-\,0.25\,{\rm \mu m}$ and a possibly high $\log\left([{\rm O\,III]\,\lambda4363/\rm H\gamma}\right)=0.17^{+0.24}_{-0.22}$ further support an AGN-like interpretation, as seen in previously known LRDs. Tentative detection of Ly$\alpha$ emission may indicate enhanced Ly$\alpha$ transmission at $z\sim10$. Our findings extend the observational frontier of BH growth into the seed-BH formation era, and upcoming MIRI spectroscopy will further establish the physical nature of the source.
\end{abstract}


\section{Introduction}\label{sec:intro}
Supermassive black holes (SMBHs) are ubiquitous in massive galaxies in the local Universe and are thought to co-evolve with their host galaxies \citep[e.g.,][]{Kormendy2013}.
However, the origin of SMBHs remains one of the major unsolved problems in modern astrophysics \citep[e.g.,][]{Volonteri2010, inayoshiAssembly2020, Regan2024_review}.

As a possible key population for uncovering the origin of SMBHs, ``little red dots'' (LRDs) have recently attracted significant attention (e.g., \citealt{Labbe2023a, Furtak2024, Matthee2024, Kocevski2023, Harikane2023_agn, Barro2024, Akins2024, Hainline2025, Hviding2025, InayoshiHo2025_review}).
LRDs were first revealed by \JWST\ thanks to its unprecedented sensitivity at infrared wavelengths \citep{Rigby2023_JWST} and are characterized by rest-optical compact morphologies and a unique V-shaped spectral energy distribution (SED), with a blue rest-UV component and a steep, red rest-optical component \citep[e.g.,][]{Labbe2023a, Furtak2024, Matthee2024, Kocevski2023, Barro2024, Akins2024}.
These components transition near the Balmer limit \citep[e.g.,][]{Setton2024} and often show a Balmer break \citep[e.g.,][]{Kokorev2024_break, Labbe2024, Wang2024a}, and in some cases exhibit an extremely strong Balmer break that cannot be explained by stellar populations alone \citep[e.g.,][]{deGraaff2025, Naidu2025, Taylor2025}.

Spectroscopic observations have revealed broad Balmer emission lines with absorption features in many LRDs \citep[e.g.,][]{Labbe2023b, Furtak2024, Matthee2024, Greene2024}.
Despite having broad Balmer lines similar to those of type-I active galactic nuclei (AGN), LRDs show several unusual properties that distinguish them from typical AGNs.
First, most LRDs are undetected in X-rays \citep{Yue2024, Ananna2024, Maiolino2024_chandra, Akins2024}, with rare exceptions including objects proposed to be in transition from LRDs to conventional AGNs \citep{Hviding2026, Fu2026}.
LRDs are also generally undetected in radio \citep{Akins2024, Perger2024, mazzolari2024, Gloudemans2025}, although radio emission has been detected in some low-$z$ LRD analogs \citep{Rodriguez2026, Rodriguez2026_2}.
Recent studies have revealed hot-dust emission in individual LRDs and stacked samples
\citep{Delvecchio2025,Lin2025_lowz,Ji2026_lord,Perez-Gonzalez2026_stack,Brazzini2026}.
However, the detection of hot-dust emission does not necessarily imply that dust attenuation is responsible for the steep red rest-optical continua of LRDs, and the relative contributions of dust attenuation and other emission mechanisms remain uncertain
\citep{Delvecchio2025,Setton2025}.
Many LRDs have not shown significant photometric variability \citep{Kokubo2024, Zhang2024_variability, tee2024_variability, Zhou2025_variability}, although photometric variability has been reported in gravitationally lensed high-redshift LRDs in a century time-scale \citep{Zhang2025_EinsteinCross} and low-$z$ analogs \citep{LinR2026_variability}.
Multi-epoch observations have also revealed variations in broad Balmer-line equivalent widths and emission-line fluxes in individual LRDs \citep{Furtak2025_variability, Ji2025, Lambrides2026_variability, Brooks2026_variability, Naidu2025}, whereas recent systematic monitoring studies have found substantially weaker H$\alpha$ variability in LRDs than expected for typical AGNs \citep{LiuZ2026_variability, Stone2026_variability}.
In summary, these unusual properties leave their physical nature uncertain.

\begin{figure*}
	\centering
	\includegraphics[width=0.98\linewidth]{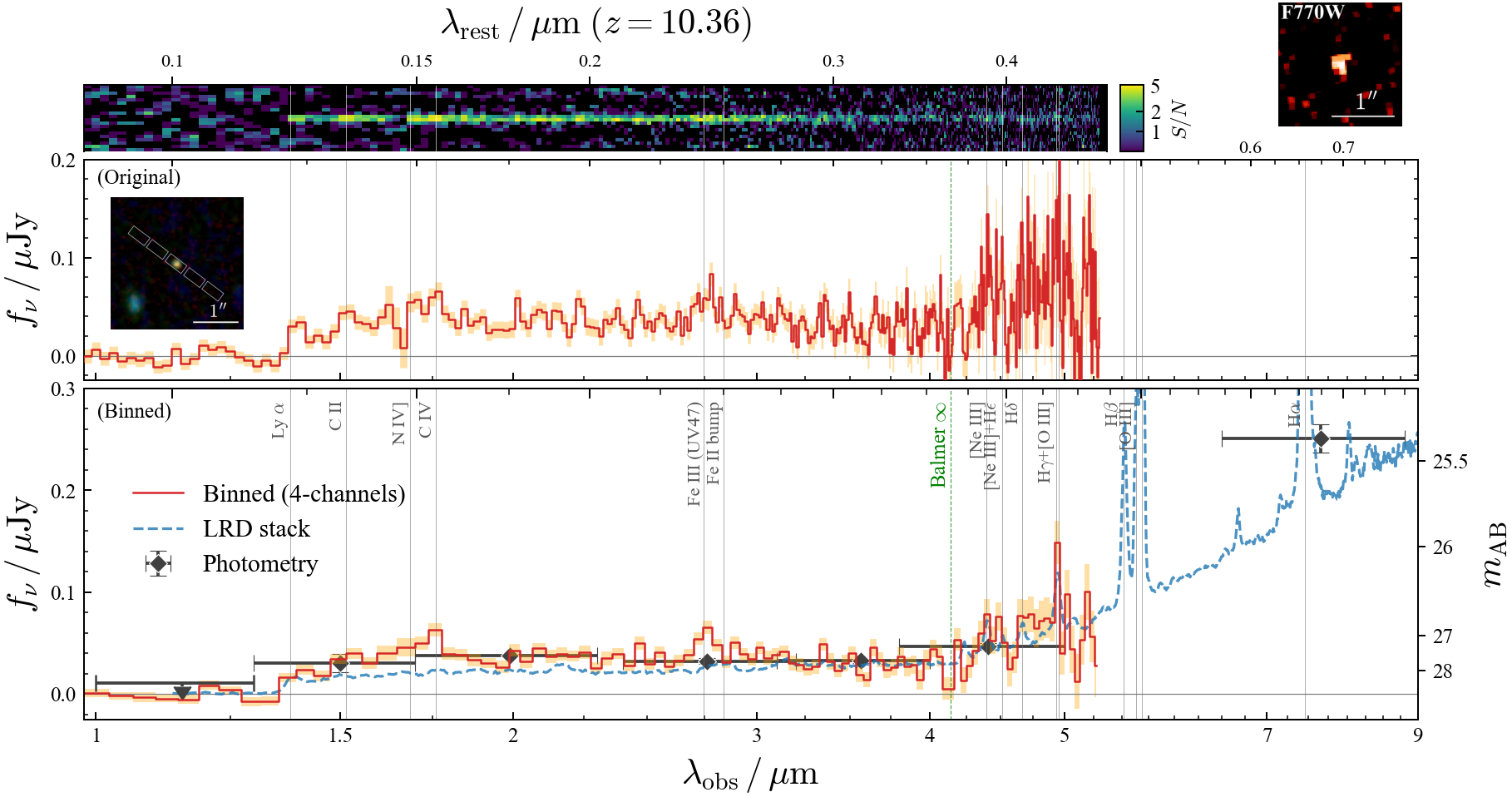}
	\caption{
    \JWST/NIRSpec PRISM spectrum of \target.
    The top panel shows the nod-subtracted 2D spectrum produced by the \JWST\ science calibration pipeline.
    The middle panel shows the optimally extracted 1D spectrum, and the bottom panel compares the spectrum, binned by taking the inverse-variance-weighted average over every four spectral channels, with the photometry reported in \cite{Tanaka2025_z10}.
    Uncertainties ($1\sigma$) are shown as the orange shaded region. 
    The inset in the middle panel shows the NIRCam $3^{\prime\prime}\times3^{\prime\prime}$ RGB (R: F356W+F444W; G: F200W+F277W; B: F115W+F150W) image overlaid with the NIRSpec slitlet, while the upper-right inset shows a $2^{\prime\prime}\times2^{\prime\prime}$ F770W image.
    In the bottom panel, we overplot a stacked spectrum of $274$ LRDs compiled from the EMBER survey and archival observations with the dashed line (H.~B.~Akins et al. \textit{in prep.}), scaled to match the continuum of \target\ over $\lambda_{\rm rest}=0.28\,\mathchar`-\,0.32\,{\rm \mu m}$.
    The spectrum exhibits the Ly$\alpha$ break at $\sim1.3\,\mathchar`-\,1.4\,{\rm \mu m}$ and a slight upturn redward of $\sim 4.2\,{\rm \mu m}$, consistent with a $V$-shaped SED of typical LRDs redshifted to $z\sim10$.}
    \label{fig:spectrum}
\end{figure*}

One proposed framework is the BH envelope (or BH-star) model \citep[e.g.,][]{Naidu2025, Kido2025, Inayoshi2025_seduni}, in which a BH rapidly grows from a seed, possibly through super-Eddington accretion, while surrounded by dense, optically thick gas with an effective temperature of $\sim4000\,\mathchar`-\,5000\,{\rm K}$.
The rest-optical-to-near-infrared SEDs of LRDs can be reproduced by blackbody-like emission from BH envelopes, whose Wien tails explain the steep rest-optical continua \citep[e.g.,][]{Lin2025_lowz, Kido2025, deGraaff2025_spec_stat, Umeda2025_BHstar}.
Furthermore, dense, predominantly neutral gas ($n_{\rm H}\gtrsim10^9\,{\rm cm^{-3}}$) can produce strong Balmer absorption and a Balmer break through collisional excitation of hydrogen to the $n=2$ state \citep{InayoshiMaiolino2024, Yan2026_Balmer}.
However, the geometry and physical properties of the surrounding gas remain uncertain, and the gas may have a covering factor below unity, as in some clumpy, disk-like, or toroidal configurations \citep{Pang2026_structure, Tang2026_SPURS, Ji2026_hole, Chen2026_ABCD, Zhou2026_SS, Yanagisawa2026}.
In addition, dense ionized gas with electron column densities of $N_{\rm e}\sim10^{24}\,{\rm cm^{-2}}$ may contribute to broad Balmer-line profiles through electron scattering \citep{Rusakov2025}, suggesting that the emitting, absorbing, and scattering regions may not be identical.
Other models invoke nebular emission from ionized gas cocoons
\citep{Sneppen2026,Sneppen2026_Pjump}, dense accretion flows around direct-collapse BHs \citep{Pacucci2026_DCBH}, or dust-reddened AGN emission \citep{Madau2026_dust,Madau2026_polarlization} to explain the characteristic red rest-optical continua of LRDs.
Therefore, we have not yet achieved a unified model of the nature of LRDs.

\begin{figure*}
	\centering
	\includegraphics[width=\linewidth]{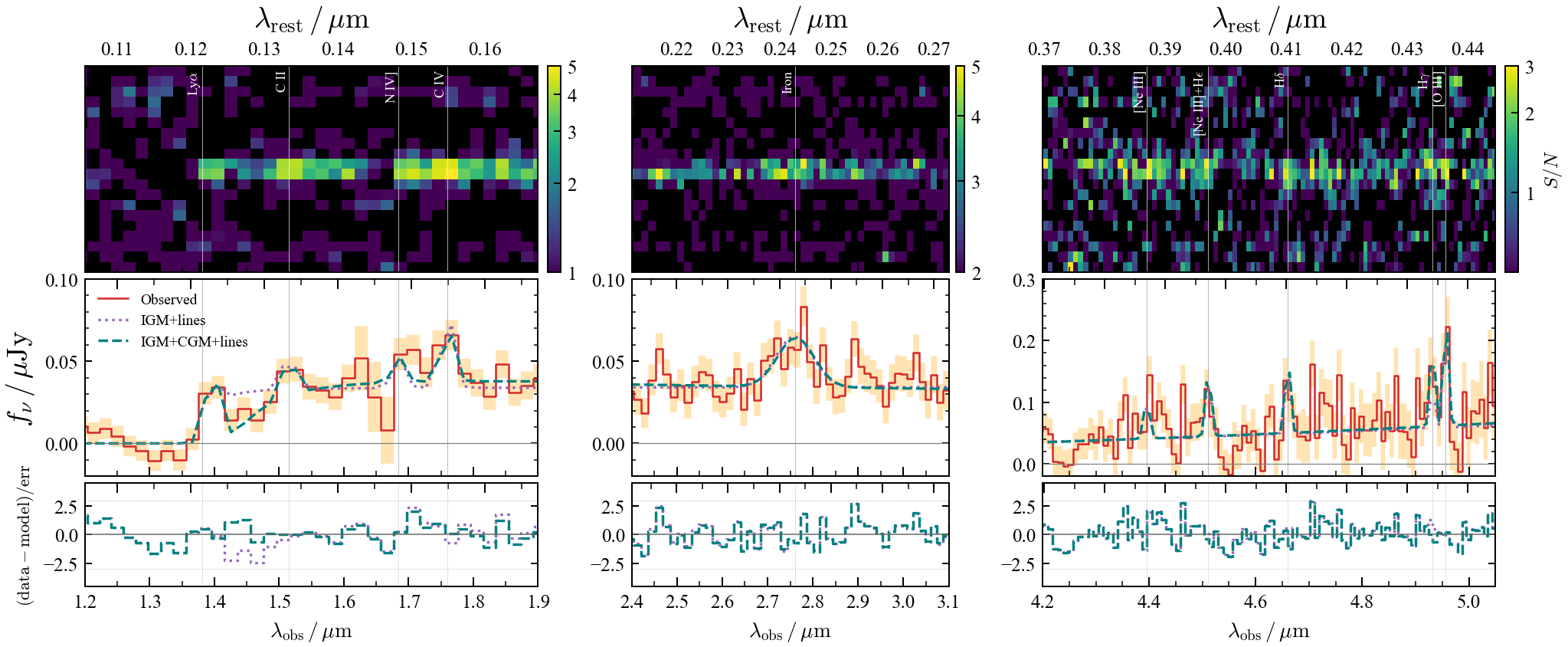}
	\caption{Zoomed-in versions of Figure\,\ref{fig:spectrum} including emission lines.
    The middle rows show the unbinned original resolution spectrum and two fitted model SEDs: IGM+lines (dotted) and IGM+CGM+lines (dashed).
    The bottom panels show the residuals normalized by the $1\sigma$ uncertainties.
    We identify multiple line features with $S/N\sim2\,\mathchar`-\,4$ at wavelengths consistent with $z_{\rm spec} = 10.36$.
    }
    \label{fig:zoomin}
\end{figure*}

Regardless of the specific physical framework invoked to explain their unusual properties, LRDs may trace an early stage of BH growth, potentially close to the epoch of BH seeding.
This picture is supported by \citet{inayoshi2025}, who compared the number density evolution of LRDs with that of non-LRD AGNs and suggested that LRDs may represent the first episode of BH growth after seed BH formation.
Therefore, LRDs provide a crucial population for observationally probing seed BHs, which have long been a theoretical target of study \citep{Volonteri2010, inayoshiAssembly2020, Regan2024_review}. 
However, almost all previous LRD studies have been limited to $z<9$.
Prior to this work, the most distant LRD with a spectroscopic redshift was CAPERS-LRD-z9 at $z=9.288$ with a broad H$\beta$ detection \citep{Taylor2025}.
More recently, \citet{Cai2026_BLAGN} reported a broad-line AGN at $z=9.984$ and discussed its possible classification as an LRD.
Extending LRD studies to $z>10$ is crucial for constraining their number-density evolution during their emergence and for uncovering when and through which mechanisms seed BHs formed, a regime that has remained observationally inaccessible until now.

As shown in Figure 1 of \cite{Tanaka2025_z10}, searching for $z > 10$ LRDs requires the mid-infrared coverage provided by MIRI to sample their steep rest-optical slope and distinguish LRDs from galaxies.
Motivated by this, \citet{Tanaka2025_z10} developed a joint NIRCam-MIRI color selection method and applied it to the COSMOS-Web imaging data (GO\#1727, PIs J.~Kartaltepe and C.~M.~Casey; \citealt{Casey2023}) and identified a promising candidate for the first $z>10$ LRD, \target, with a photometric redshift of $z_{\rm phot}=10.5^{+0.7}_{-0.6}$.
They used this candidate to place the first constraint on the abundance of LRDs at $z\sim10$, $\Phi_{\rm LRD}=\left(1.2^{+2.7}_{-1.0}\right)\times10^{-6}\,{\rm Mpc^{-3}\,mag^{-1}}$ for $M_{\rm UV}=-20$, suggesting the redshift evolution of their number density following the log-normal evolution hypothesized by \citet{inayoshi2025}.
However, \target\ had not yet been spectroscopically confirmed, leaving both its nature and the inferred $z\sim10$ LRD abundance dependent on the photometric interpretation.

In this paper, we report NIRSpec spectroscopic observations of \target.
In Section~\ref{sec:data}, we describe the data.
In Sections~\ref{sec:spec_fitting} and \ref{sec:results}, we present the spectroscopic analysis and the results, respectively.
Then Section\,\ref{sec:discussion} discusses the nature of \target, and Section\,\ref{sec:summary} provides the summary.
Throughout this work, we assume a flat $\Lambda$CDM cosmology with $H_0 = 67.4\,{\rm km\,s^{-1}\,Mpc^{-1}}$, $\Omega_{\rm m} = 0.315$, and $\Omega_\Lambda=0.685$ \citep{Planck2018}.

\section{Data}\label{sec:data}
\target\ at $(\mathrm{R.A.},\mathrm{Decl.})=(10^{\rm h}00^{\rm m}20^{\rm s}.1836,$ $+02^\circ27^{\prime}14.\!^{\prime\prime}022)$ was observed with \JWST/NIRSpec Multi-Object Spectroscopy as part of the Early Massive Black Hole Evolution \& Reddening (EMBER) program (GO\#7076, PI H.~Akins).
EMBER was designed to conduct a comprehensive spectroscopic census of the brightest LRDs selected across multiple extragalactic fields. 
While the highest priority EMBER targets were bright ($m_{\rm F444W}<24$) $z\sim 3$--$7$ LRDs, \target\ was selected for observation based on its robust photometric redshift of $z_{\rm phot}=10.5^{+0.7}_{-0.6}$ and previous identification as an LRD candidate based on MIRI photometry \citep{Tanaka2025_z10}. 
\target\ was observed as part of Observation\,1 of the EMBER program on November 27, 2025, with on-source exposure times of 1.48\,hr using the PRISM/CLEAR disperser/filter pair ($0.6\,\mathchar`-\,5.3\,{\rm \mu m}$, $R\sim100$) and 0.92\,hr using G395M/F290LP ($2.87\,\mathchar`-\,5.10\,{\rm \mu m}$, $R\sim1000$). 

We process the NIRSpec data starting from the uncalibrated \texttt{uncal} products using version 1.20.2 of the \JWST\ Calibration Pipeline \citep{bushouseJWST2023} and the Calibration Reference Data System (CRDS) mapping pmap-1481. 
The reduction follows the standard workflow implemented by the pipeline, with a few modifications to improve the data quality (described in more detail in H.~B.~Akins et al.~\textit{in prep}). 
First, following the detector-level corrections applied in stage~1, we employ a custom iterative procedure to subtract the thermal ``picture frame'' effect and $1/f$ noise on the detector, masking spectral traces from all open micro-shutter assembly (MSA) shutters. 
The picture frame subtraction uses the recently released CRDS \verb|pictureframe| reference file, and is particularly important for reliable background subtraction. 
We extract the 2D spectra for \target\ from each nod in stage~2 and apply flat-field, photometric calibration, and wavelength calibration. 
We inspect each nod to ensure that no shutters were stuck closed, and then perform the standard nodded background subtraction and combination of the individual exposures into rectified products. 
We extract a 1D spectrum from the combined 2D product using the optimal-extraction method of \cite{horneOptimal1986} with a profile constructed as the median of the rectified 2D spectrum along the cross-dispersion direction. 
We note that the G395M spectrum for \target\ is contaminated by the spectral trace from a nearby bright object, as G395M spectra were allowed to overlap in the MSA design.
Given the contamination and the faint nature of this source, the analysis here focuses only on the deeper PRISM spectrum. 
We present the processed NIRSpec/PRISM spectrum in Figure\,\ref{fig:spectrum}.
The spectrum is consistent with the observed photometry, supporting the robustness of the pipeline slit-loss correction.

\begin{deluxetable*}{ccc|cccccccc}
\tabletypesize{\scriptsize}
\tablecaption{Estimated properties of \target\ from the spectral fits.
\label{tab:fit_model}}
\tablewidth{0pt}
\tablehead{
\multicolumn{3}{c}{Model} &
\colhead{$z_{\rm spec}$} &
\colhead{$M_{\rm UV}$} &
\colhead{$\beta_{\rm UV}$} &
\colhead{$\beta_{\rm opt}$} &
\colhead{$\lambda_{\rm turn}$} &
\colhead{$\log R_{\rm b}$} &
\colhead{$\log N_{\rm HI}$} &
\colhead{$\Delta {\rm BIC}^{*}$} \\
\colhead{Continuum} &
\colhead{Absorption} &
\colhead{Lines$^\dag$} &
\colhead{} &
\colhead{} &
\colhead{} &
\colhead{} &
\colhead{[$\mu$m]} &
\colhead{[cMpc]} &
\colhead{[cm$^{-2}$]} &
\colhead{}
}
\startdata
BPL & IGM+CGM & Included &
$10.361^{+0.003}_{-0.002}$ &
$-20.16^{+0.07}_{-0.07}$ &
$-2.26^{+0.11}_{-0.11}$ &
$1.17^{+0.74}_{-0.70}$ &
$0.356^{+0.010}_{-0.013}$ &
$0.42^{+1.04}_{-0.97}$ &
$22.81^{+0.20}_{-0.25}$ &
--- \\
BPL & IGM & Included &
$10.361^{+0.003}_{-0.002}$ &
$-20.01^{+0.05}_{-0.05}$ &
$-2.03^{+0.08}_{-0.08}$ &
$1.01^{+0.71}_{-0.69}$ &
$0.359^{+0.011}_{-0.013}$ &
$0.65^{+0.63}_{-1.10}$ &
--- &
2.6 \\
BPL & IGM+CGM & Ly$\alpha$ only &
$10.41^{+0.06}_{-0.06}$ &
$-20.13^{+0.07}_{-0.08}$ &
$-2.20^{+0.12}_{-0.12}$ &
$1.09^{+0.77}_{-0.70}$ &
$0.355^{+0.012}_{-0.014}$ &
$0.56^{+1.10}_{-1.06}$ &
$22.52^{+0.31}_{-0.53}$ &
--- \\
BPL & IGM & Ly$\alpha$ only &
$10.32^{+0.16}_{-0.09}$ &
$-20.00^{+0.05}_{-0.06}$ &
$-2.03^{+0.08}_{-0.09}$ &
$1.02^{+0.80}_{-0.75}$ &
$0.362^{+0.013}_{-0.016}$ &
$0.38^{+1.08}_{-0.95}$ &
--- &
--- \\
SPL & IGM+CGM & Included &
$10.361^{+0.002}_{-0.002}$ &
$-20.05^{+0.07}_{-0.07}$ &
$-2.01^{+0.11}_{-0.10}$ &
--- &
--- &
$0.46^{+0.96}_{-0.97}$ &
$22.56^{+0.28}_{-0.42}$ &
17.7 \\
\enddata
\tablenotetext{\dag}{
(Included) The following emission lines are modeled with Gaussian profiles: Ly$\alpha$, C\,{\sc ii}\,$\lambda1335$, N\,{\sc iv}]\,$\lambda\lambda1483,1487$, C\,{\sc iv}\,$\lambda\lambda1549,1551$, an iron complex (Fe\,{\sc ii} UV bump or Fe\,{\sc iii} UV47 multiplet), [Ne\,{\sc iii}]\,$\lambda3869$, [Ne\,{\sc iii}]\,$\lambda3967$ + H$\epsilon$, H$\delta$, H$\gamma$, and [O\,{\sc iii}]\,$\lambda4363$.
(Ly$\alpha$ only) Only Ly$\alpha$ is included in the model, and the corresponding wavelength regions for the other lines are masked.
}
\tablenotetext{*}{
BIC difference compared to the broken power-law model with the IGM+CGM absorption and the lines.
The BIC differences are not calculated for models without emission lines, as the corresponding wavelength regions are masked during fitting.
}
\end{deluxetable*}

\begin{deluxetable*}{llll}
\tablecaption{
Estimated line fluxes and equivalent widths of \target\ from the fitting with the BPL model including IGM+CGM absorption and emission lines.
\label{tab:line}
}
\tablewidth{0pt}
\tablehead{
\colhead{Line} &
\colhead{$F_{\rm line}/10^{-20}$} &
\colhead{$S/N_{\rm line}$} &
\colhead{${\rm EW}_{\rm rest}$} \\
&
\colhead{[${\rm erg\,s^{-1}\,cm^{-2}}$]} &
&
\colhead{[\AA]}
}
\startdata
Ly$\alpha$ &
$243^{+40}_{-51}$ &
4.2 &
$34.1^{+6.6}_{-7.3}$ \\
C\,{\sc ii}\,$\lambda1335$ &
$73.5^{+30.6}_{-29.2}$ &
2.5 &
$12.4^{+5.2}_{-5.1}$ \\
{N\,{\sc iv}]}\,$\lambda\lambda1483,1487$ &
$46.9^{+26.8}_{-25.2}$ &
1.9 &
$10.0^{+6.0}_{-5.4}$ \\
C\,{\sc iv}\,$\lambda\lambda1549,1551$ &
$94.3^{+23.7}_{-24.4}$ &
3.9 &
$22.2^{+6.1}_{-5.9}$ \\
Iron complex &
$121.4^{+27.5}_{-27.2}$ &
4.4 &
$79.2^{+18.8}_{-18.7}$ \\
{[Ne\,{\sc iii}]}\,$\lambda3869$ &
$14.8^{+6.9}_{-6.3}$ &
2.2 &
$20.7^{+9.7}_{-9.1}$ \\
{[Ne\,{\sc iii}]}\,$\lambda3967$ + H$\epsilon$ &
$14.9^{+6.3}_{-6.1}$ &
2.4 &
$20.0^{+8.9}_{-8.5}$ \\
H$\delta$ &
$16.6^{+6.2}_{-6.8}$ &
2.6 &
$21.6^{+8.6}_{-9.0}$ \\
H$\gamma$ &
$17.8^{+7.4}_{-7.1}$ &
2.5 &
$21.4^{+10.2}_{-8.6}$ \\
{[O\,{\sc iii}]}\,$\lambda4363$ &
$26.5^{+7.0}_{-7.8}$ &
3.6 &
$31.8^{+9.7}_{-9.8}$ \\
\enddata

\tablecomments{
The quoted uncertainties correspond to the 16th--84th percentile
ranges of the posterior distributions.
All lines except the iron complex and Ly$\alpha$ are assumed to be unresolved, with their observed widths determined by the wavelength-dependent NIRSpec/PRISM line-spread function.
The iron complex is treated as a resolved feature, with a fitted effective rest-frame central wavelength of $0.2432^{+0.0011}_{-0.0010}\,{\rm \mu m}$ and an intrinsic FWHM of $1.22^{+0.58}_{-0.32}\times10^4\,{\rm km\,s^{-1}}$.
}
\end{deluxetable*}

\section{Spectroscopic fitting}\label{sec:spec_fitting}

To derive the spectroscopic redshift and characterize \target, we perform Markov Chain Monte Carlo spectrum fitting using {\tt emcee} \citep{emcee}.
For the intrinsic continuum, we adopt a broken power law (BPL) with independent rest-frame UV and optical slopes, $\beta_{\rm UV}$ and $\beta_{\rm opt}$ ($f_\lambda \propto \lambda^{\beta}$).
The two power laws are joined continuously at a turnover rest-frame wavelength, $\lambda_{\rm turn}$, and the overall normalization is parameterized by the absolute magnitude in rest-frame 1450\,{\AA} ($M_{\rm UV}$).
To test whether the BPL model is required, we additionally fit a single power-law (SPL) continuum and compare the goodness of fit using the Bayesian information criterion (BIC).

We include Gaussian emission lines for Ly$\alpha$, C\,{\sc ii}\,$\lambda1335$, N\,{\sc iv}]\,$\lambda\lambda1483,1487$, C\,{\sc iv}\,$\lambda\lambda1549,1551$, [Ne\,{\sc iii}]\,$\lambda3869$, [Ne\,{\sc iii}]\,$\lambda3967$ + H$\epsilon$, H$\delta$, H$\gamma$, and [O\,{\sc iii}]\,$\lambda4363$.
Because these lines are unresolved at the spectral resolution of NIRSpec/PRISM, their observed profiles are modeled using a Gaussian line-spread function (LSF) with a wavelength-dependent resolution for a point source provided by the \texttt{unite} package ($R\sim50\,\mathchar`-\,500$; \citealt{unite}).
This prescription assumes a point source centered in an MSA shutter, which is a reasonable approximation for the observation of \target\ (Figure\,\ref{fig:spectrum}), and accounts for the higher effective spectral resolution of compact sources compared to the nominal resolution derived assuming uniform illumination across the shutter (e.g., see \citealt{deGraaff2024_msa}).
Note that, at the fitted spectroscopic redshift of $z_{\rm spec}=10.36$ (Section\,\ref{ssec:spec_z}), [O\,{\sc iii}]\,$\lambda4363$ and H$\gamma$ are observed near $\lambda_{\rm obs}\simeq5\,\mu{\rm m}$, where the PRISM resolving power reaches $R\sim450$ based on the \texttt{unite} LSF, corresponding to a rest-frame spectral resolution of $\Delta\lambda_{\rm rest}\sim10\,\text{\AA}$. 
Therefore, [O\,{\sc iii}]\,$\lambda4363$ and H$\gamma$, separated by $\sim23\,\text{\AA}$ in the rest frame, are spectrally distinguishable in principle.

Given the limited spectral resolution of PRISM, we fit the unresolved N\,{\sc iv}]\,$\lambda\lambda1483,1487$ and C\,{\sc iv}\,$\lambda\lambda1549,1551$ doublets with a single Gaussian profile each, and tie the centroids of all lines except Ly$\alpha$ and the iron complex to a common systemic redshift without allowing individual velocity offsets.
Thus, the only free parameter for each of these unresolved lines is its line flux.
Note that the iron complex, which likely consists of a blend of multiple iron emission lines, is treated separately, and both its intrinsic FWHM and effective rest-frame central wavelength are fitted as free parameters.

Ly$\alpha$ is also treated separately, and both its intrinsic FWHM and velocity offset are fitted as free parameters.
We consider two model setups: (i) only intergalactic medium (IGM) absorption and (ii) both IGM and circumgalactic medium (CGM) Ly$\alpha$ absorption.
IGM absorption is modeled following the formalism of \cite{Totani2006}, which is based on \cite{Miralda1998}, with two free parameters: the neutral hydrogen fraction $x_{\rm HI}$ and the ionized bubble radius $R_{\rm b}$.
Note that, following \cite{Umeda2024_IGM}, we fix the redshift corresponding to the end of reionization at $z_{\rm end}=5$ and neglect the contribution of neutral IGM at $z<z_{\rm end}$ to the damping-wing absorption. 
We confirm that varying $z_{\rm end}$ between 5 and 6 has a negligible effect on the fitting results.
For the CGM, we assume a Voigt absorption profile with three free parameters: the neutral hydrogen column density $\log\left(N_{\rm H I}\right)$, the velocity offset of the CGM from the galaxy $\Delta v$, and the velocity dispersion of the CGM $\sigma_{\rm CGM}$.

We also fit an emission feature that possibly corresponds to an iron complex around $\lambda_{\rm rest}\sim0.24\,\mathchar`-\,0.25\,{\rm \mu m}$ (see also Section\,\ref{ssec:iron}).
Since the individual lines are blended and cannot be resolved, we model the entire feature with a single Gaussian profile, treating its FWHM and central wavelength as free parameters.
We confirm that the overall feature can be adequately reproduced by a single broad Gaussian with ${\rm FWHM}\simeq1.22^{+0.58}_{-0.32}\times10^4\,{\rm km\,s^{-1}}$ at the central wavelength of $\lambda_{\rm rest}=0.2432^{+0.0011}_{-0.0010}\,{\rm \mu m}$.
We emphasize that the primary purpose of this fit is to assess the significance of the iron-complex detection by using the integrated flux derived from the fitted Gaussian component, and the fitted line width is not assigned any physical interpretation.
The best-fit SEDs are shown in Figure\,\ref{fig:zoomin}, and the fitted parameters are shown in Tables\,\ref{tab:fit_model} and \ref{tab:line}. 
\section{Results}\label{sec:results}

\subsection{Spectroscopic redshift}\label{ssec:spec_z}
The derived redshift is $z_{\rm spec} = 10.361^{+0.003}_{-0.002}$ from both the IGM+lines and the IGM+CGM+lines models, consistent with each other and with the previous photometric redshift, $z_{\rm phot}=10.5^{+0.7}_{-0.6}$ \citep{Tanaka2025_z10}.
These tight constraints are largely driven by the tentative line features discussed in Section\,\ref{ssec:lines}.

Independently of these lines, the spectrum also exhibits a clear break feature around $\lambda_{\rm obs}\sim 1.4\,{\rm \mu m}$, which corresponds to the Ly$\alpha$ break.
\target\ is undetected blueward of the break, while the continuum is detected at a signal-to-noise ratio of $S/N\sim4\,\mathchar`-\,5$ redward of the break.
To assess the redshift constraint primarily from the continuum break, we repeat the fit after masking the wavelength regions of the tentative line detections. 
Note that Ly$\alpha$ is left unmasked and remains included in the model, because masking the Ly$\alpha$ region would also remove the Ly$\alpha$ break.
As shown in Table\,\ref{tab:fit_model}, the derived redshifts from the IGM and the IGM+CGM models, $10.32^{+0.16}_{-0.09}$ and $z_{\rm spec}=10.41^{+0.06}_{-0.06}$, respectively, are consistent with $z_{\rm spec}$ obtained including the lines.

Note that this break cannot be interpreted as a Balmer break at $z\sim3$.
When combined with the non-detections in filters blueward of $1.4\,{\rm \mu m}$, the implied break strength, defined as the $f_\nu$ ratio across the break, is $\gtrsim10$.
Such an extreme Balmer break is far stronger than expected for a conventional post-starburst galaxy, for which the Balmer break strength is typically $\lesssim3$.
We also confirm this by fitting the photometry with $z=3\,\mathchar`-\,4$ galaxy models using \texttt{Prospector} \citep{Johnson2021}.
The best-fit model SED predicts a continuum flux density of $\sim0.02\,\mu{\rm Jy}$ at $\lambda_{\rm obs}<1.5\,\mu{\rm m}$, which is inconsistent with both the observed spectrum and the photometric upper limits, e.g., F115W $<3.3\,{\rm nJy}$ \citep[$2\sigma$ upper limit by][]{Tanaka2025_z10}.
Although similarly extreme Balmer breaks have been reported in some LRDs \citep{Naidu2025, deGraaff2025, Taylor2025}, an LRD-like solution at $z\sim3$ is also disfavored by the absence of strong rest-optical emission lines, such as H$\alpha$.
Furthermore, the continuum slope at $\lambda_{\rm obs}>1.5\,\mu{\rm m}$, $\beta=-2.01^{+0.11}_{-0.10}$ from the SPL fit, is substantially bluer than the rest-optical slopes expected for LRDs, for example $\beta_{\rm opt}>0$ \citep{Kocevski2024} or $\beta_{\rm opt}>-0.52$ \citep{Kocevski2026}.

\begin{figure}
	\centering
	\includegraphics[width=\linewidth]{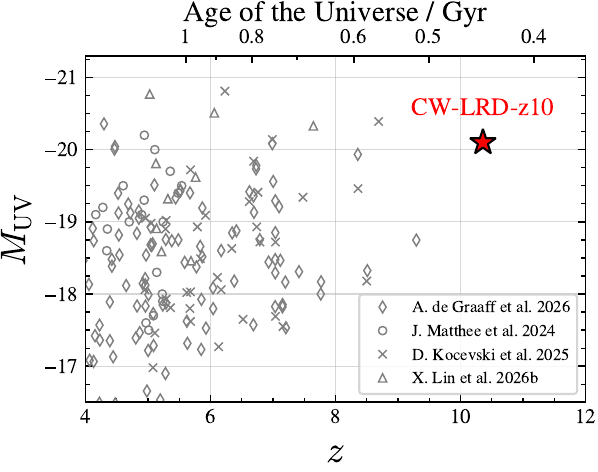}
	\caption{UV absolute magnitudes ($M_{\rm UV}$) of previously spectroscopically confirmed LRDs \citep{Matthee2024,Kocevski2024,Lin2025_c3d,deGraaff2025_spec_stat} as a function of redshift.
    Note that there can be an overlap between different studies.
    The red star indicates \target\ (this study), representing the first LRD candidate with $z_{\rm spec}>10$, corresponding to a cosmic age of $<0.5\,{\rm Gyr}$.
    }\label{fig:muv_z}
\end{figure}

Therefore, we conclude that \target\ is at $z_{\rm spec}=10.36$, which establishes \target\ as the first $z_{\rm spec}>10$ LRD candidate.
The obtained UV absolute magnitude of $M_{\rm UV}=-20.16^{+0.07}_{-0.07}$ places it among the most luminous LRDs (Figure\,\ref{fig:muv_z}).
It is substantially brighter than the only previously published LRD at $z_{\rm spec}>9$, CAPERS-LRD-z9, at $z=9.288$ and $M_{\rm UV}=-18.2\pm0.2$ \citep{Taylor2025}.

\begin{figure*}
	\centering
	\includegraphics[width=0.9\linewidth]{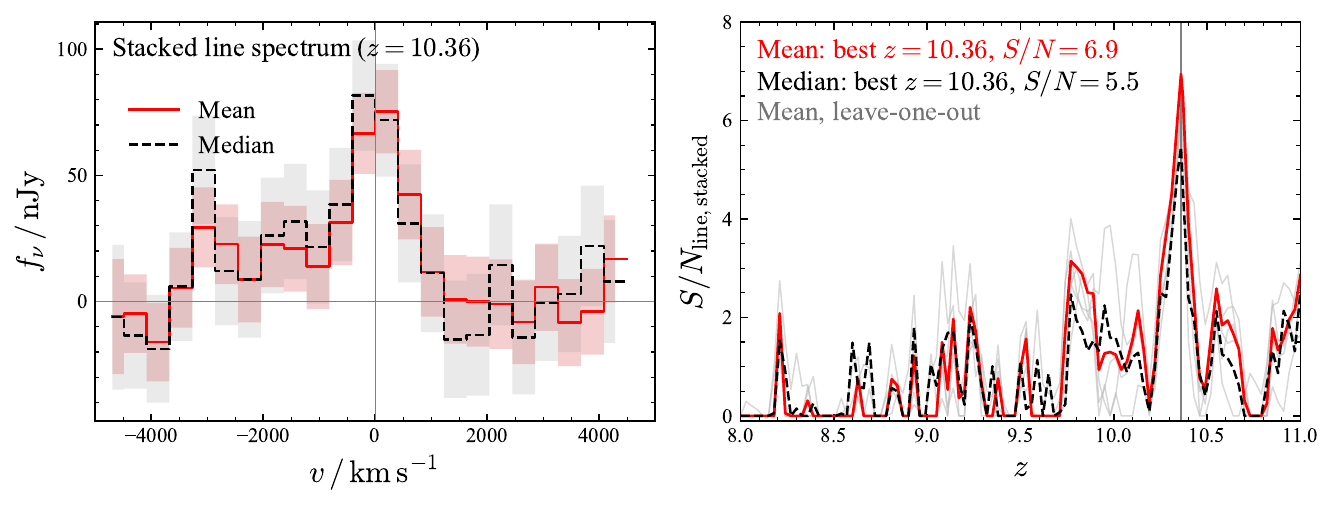}
	\caption{(Left) Continuum-subtracted stacked spectrum of [Ne\,{\sc iii}]\,$\lambda3869$, [Ne\,{\sc iii}]\,$\lambda3967$ + H$\epsilon$, H$\delta$, H$\gamma$, and [O\,{\sc iii}]\,$\lambda4363$ assuming $z=10.36$. 
    The red solid and black dashed lines show mean- and median-stacked spectra, respectively.
    $1\sigma$ uncertainties are shown as the shaded regions.
    (Right) $S/N$ from the Gaussian fitting when redshift is varied. 
    The gray thin lines indicate the results from the leave-one-out test.
    The stack $S/N$ peaks at $z\sim10.36$, providing additional support for $z_{\rm spec}=10.36$.}
    \label{fig:stack}
\end{figure*}

\subsection{Spectroscopic confirmation of a V-shaped SED}\label{ssec:v_shape}
The fit with the BPL continuum with IGM+CGM absorption and emission lines results in $\beta_{\rm UV}=-2.26^{+0.11}_{-0.11}$ and $\beta_{\rm opt}=1.17^{+0.74}_{-0.70}$.
Both of the fitted spectral slopes, as well as their large difference across the turnover, are consistent with those typically observed in lower-$z$ LRDs \citep[e.g.,][]{Kocevski2024, deGraaff2025_spec_stat, Barro2025, Ando2026_UV}.
Furthermore, the BPL model is strongly favored over the SPL model with $\beta=-2.01^{+0.11}_{-0.10}$ by the BIC improvement of $\Delta{\rm BIC}=17.7>10$, a conventional threshold for strong evidence in model selection, confirming a significant change in the continuum slope.
This continuum-slope change is more clearly visible in the 4-channel binned spectrum (Figure\,\ref{fig:spectrum}, lower).
Furthermore, the fitted turnover wavelength is $\lambda_{\rm turn}=0.356^{+0.010}_{-0.013}\,{\rm \mu m}$, which is consistent with the Balmer limit ($\lambda_{\rm rest}=0.3646\,{\rm \mu m}$) within the uncertainties, as commonly seen in LRDs \citep[e.g.,][]{Setton2024,deGraaff2025_spec_stat}.

The UV-to-optical continuum shape and characteristic V-shape turnover in \target\ are consistent with typical LRDs at lower redshift. 
In the bottom panel of Figure\,\ref{fig:spectrum}, we overplot a stack of PRISM spectra of 274 LRDs from $z\sim 2.3$--$9.3$, combining the EMBER survey with PRISM spectra from the ZENITH program (GO\#7417, PIs C.~Casey, H.~B.~Akins, and M.~Franco) and archival observations from CAPERS (GO\#6368, \citealt{Taylor2025}), RUBIES (GO\#4233, \citealt{deGraaff2025_RUBIES}), and other archival programs. 
The details of the LRD selection and stacking procedure will be provided in H.~B.~Akins et al.~(\textit{in prep}). 
The stacked spectrum shows a clear turnover at the Balmer limit, similar to \target, and continues to rise toward redder wavelengths, consistent with the observed F770W photometry of \target.

Moreover, \target's Balmer-break strength, defined as the ratio of the continuum flux densities across the Balmer limit, is 
\begin{equation}
    \frac{\langle f_\nu \rangle_{4000\,\mathchar`-\,4100\,\text{\AA}}}{\langle f_\nu \rangle_{3620\,\mathchar`-\,3720\,\text{\AA}}} = 1.22^{+0.45}_{-0.35}.
\end{equation}
This value is comparable to those commonly observed in LRDs at $z\sim4\,\mathchar`-\,7$ \citep{deGraaff2025_spec_stat} and the value from the stacked spectrum ($\sim 1.6$).
However, the measured ratio is consistent with unity within a $1\sigma$ uncertainty and does not provide significant evidence for a Balmer break.
Nevertheless, spectral fitting over a broader wavelength range strongly favors a broken power-law continuum, with different UV and optical slopes joined near the Balmer limit.
We therefore conclude that the characteristic V-shaped continuum is still supported by the spectrum.

\subsection{Detection of rest-UV and optical lines}\label{ssec:lines}
Based on the spectroscopic fitting, we detect multiple emission-line features at the $S/N\sim 2\,\mathchar`-\,4$ level. 
Other than Ly$\alpha$ (Section\,\ref{ssec:Lya}), C\,{\sc iv}\,$\lambda\lambda1549,1551$ and [O\,{\sc iii}]\,$\lambda4363$ show the highest values with $S/N=3.9$ and $3.6$, respectively.

To test these tentative line detections, we perform spectral stacking of these lines.
After subtracting the fitted BPL continuum, we stack the residual spectrum around five rest-optical lines, [Ne\,{\sc iii}]\,$\lambda3869$, [Ne\,{\sc iii}]\,$\lambda3967$ + H$\epsilon$, H$\delta$, H$\gamma$, and [O\,{\sc iii}]\,$\lambda4363$ (Figure\,\ref{fig:stack}\,left), in the velocity plane within $v\in[-5000, 5000]\,{\rm km\,s^{-1}}$.
In the stacking process, other lines within the extracted velocity range are masked within a $\pm1000\,{\rm km\,s^{-1}}$ window, and we construct both an inverse-variance-weighted mean stack and a median stack.
We repeat this stacking over a redshift range of $z=8\,\mathchar`-\,11$ and fit the stacked spectrum with a Gaussian centered at $v=0$ to evaluate its $S/N$.
Note that we use the stacked $S/N$ as a consistency check between each line rather than as an independent global detection significance.
We do not include the rest-UV lines shown in Figure\,\ref{fig:zoomin}\,left (Ly$\alpha$, C\,{\sc ii}\,$\lambda1335$, N\,{\sc iv}]\,$\lambda\lambda1483,1487$, and C\,{\sc iv}\,$\lambda\lambda1549,1551$) in the stacking analysis, because they have a much lower spectral resolution than optical lines. 

The stacked signals from the mean and median stacking methods reach their maximum significance at $z=10.36$, with $S/N=6.9$ and $5.5$ for mean and median, respectively (Figure\,\ref{fig:stack}).
As a robustness check, we perform a leave-one-out test by repeating the stacking analysis five times, excluding one of the five lines in each realization.
The stacked excess with $S/N=6.2\,\mathchar`-\,6.4$ at $z=10.36$ remains present in all cases, indicating that the stacking result is not driven by any single feature.
Therefore, we conclude that these tentative detections are unlikely to be coincidental and support the fitted spectroscopic redshift of $z_{\rm spec}=10.36$.

\begin{figure*}
	\centering
	\includegraphics[width=0.9\linewidth]{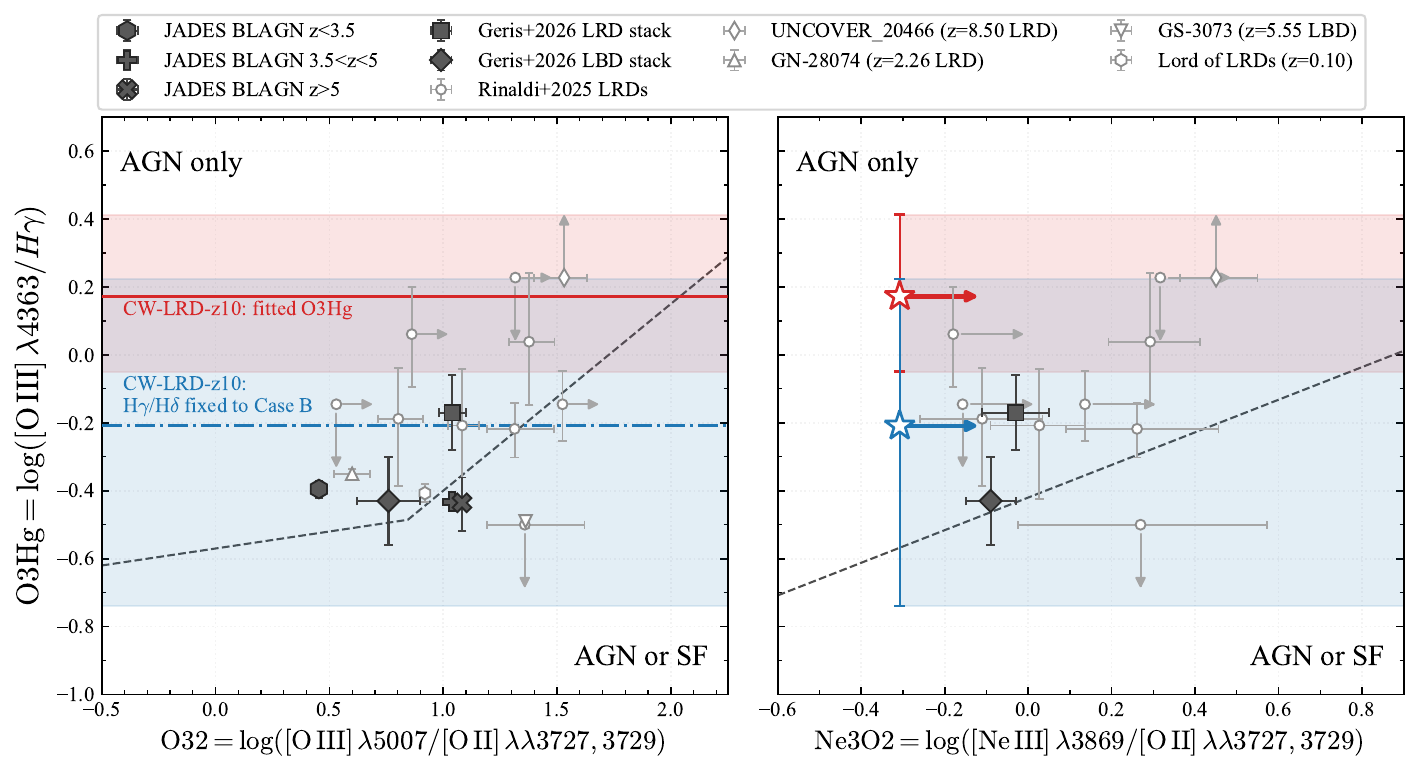}
	\caption{
    (left) O3Hg--O32 and (right) O3Hg--Ne3O2 line diagnostic diagrams.
    The black dashed line shows the empirical threshold proposed by \citet{Mazzolari2024_4363}, above which sources can be uniquely classified as AGNs (``AGN only''), while both AGNs and star-forming galaxies can occupy the region below (``AGN or SF'').
    In the left panel, the red solid and blue dot-dashed lines show the median O3Hg and the O3Hg after blending correction assuming the Case-B Balmer-line ratios between H$\gamma$ and H$\delta$.
    Note that no constraint on O32 is available because [O\,{\sc iii}]\,$\lambda5007$ falls outside the observed wavelength range.
    In the right panel, the stars and arrows indicate the conservative lower limit of Ne3O2, derived from the $3\sigma$ upper limit on the [O\,{\sc ii}]\,$\lambda\lambda3727,3729$ flux and the 16th percentile of the [Ne\,{\sc iii}]\,$\lambda3869$ flux posterior.
    $1\sigma$ confidence levels are shown as the shaded regions.
    For comparison, we show the stacked spectra of JADES broad-line AGNs \citep{Juodzbalis2025} and LRDs and LBDs \citep{Geris2026} as gray filled markers, and individual sources from the literature as light-gray open markers: UNCOVER\_20466 \citep{Jones2026_4363}, GN-28074 and GS-3073 \citep{Brazzini2026}, the Lord of LRDs \citep{Ji2026_lord}, and the LRD sample from \citet{Rinaldi2024}.
    Although the inferred line ratios are consistent with the AGN-only model and those observed in LRDs and AGNs, ionization by star formation cannot be ruled out given the uncertainties and possible line blending.
    }
    \label{fig:O3Hg_O32}
\end{figure*}

\subsection{
Line diagnostics
}\label{ssec:line_diag}
Among the line-flux ratios accessible from the detected lines, ${\rm O3Hg}\equiv\log\left({\rm [O\,III]\,\lambda4363}/{\rm H\gamma}\right)$ provides an important diagnostic of the nature of the ionizing source.
[O\,{\sc iii}]\,$\lambda4363$ is sensitive to the electron temperature and the O3Hg ratio can be enhanced by the hard ionizing radiation field produced by an AGN \citep[e.g.,][]{Gutkin2016, Feltre2016, Mazzolari2024_4363}.
Some studies \citep[e.g.,][]{Gutkin2016, Feltre2016, Mazzolari2024_4363, Martin2026} have also used line diagnostics involving O3Hg in combination with line ratios such as O33 ([O\,{\sc iii}]\,$\lambda5007$/[O\,{\sc iii}]\,$\lambda4363$), O32 ([O\,{\sc iii}]\,$\lambda5007$/[O\,{\sc ii}]\,$\lambda\lambda3727,3729$), and Ne3O2 ([Ne\,{\sc iii}]\,$\lambda3869$/[O\,{\sc ii}]\,$\lambda3727$) for discussing the nature of the ionizing source.
However, the limited depth and wavelength coverage of the available data result in a marginal detection of [Ne\,{\sc iii}]\,$\lambda3869$ ($S/N=2.2$), a non-detection of [O\,{\sc ii}]\,$\lambda\lambda3727,3729$, and no coverage of [O\,{\sc iii}]\,$\lambda5007$.

The O3Hg ratio derived from the posterior samples is ${\rm O3Hg}=0.17^{+0.24}_{-0.22}$.
This value is higher than those measured from stacked spectra of LRDs, LBDs \citep{Geris2026}, and broad-line AGNs \citep{Juodzbalis2025} observed with JWST, and is relatively high even among measurements for individual LRDs \citep{Rinaldi2024, Jones2026_4363, Brazzini2026, Ji2026_lord}.
It is consistent with the values expected from AGN-only models over a wide range of O32$\lesssim2$ and Ne3O2$\lesssim1$ (Figure\,\ref{fig:O3Hg_O32}).
We also estimate a $3\sigma$ upper limit on the [O\,{\sc ii}]\,$\lambda\lambda3727,3729$ flux using mock spectra and obtain $F_{\rm [O\,II]}<1.73\times10^{-19}\,{\rm erg\,s^{-1}\,cm^{-2}}$.
Combining this upper limit with the 16th percentile of the [Ne\,{\sc iii}]\,$\lambda3869$ flux posterior, we derive a conservative lower limit of ${\rm Ne3O2}>-0.31$.
This constraint is consistent with the AGN-only region of the diagnostic diagram over a wide range of plausible Ne3O2 values (Figure\,\ref{fig:O3Hg_O32}).
However, neither of these line diagnostics rule out ionization by star formation.

The observed ${\rm H\gamma}/{\rm H\delta}$ flux ratio ${\rm H\gamma}/{\rm H\delta}=1.07^{+0.95}_{-0.47}$ is lower than the value expected for Case B recombination, which is $\simeq 1.8$ for $T_{\rm e}=1\times10^4\,{\rm K}$ and $n_{\rm e}=1\times10^2\,{\rm cm}^{-3}$.
This discrepancy cannot be explained by dust attenuation, since dust would preferentially suppress H$\delta$ relative to H$\gamma$ and therefore increase the ${\rm H\gamma}/{\rm H\delta}$ ratio.
Previous studies have discussed that Balmer-line ratios in the dense gas associated with LRDs may deviate from the Case B recombination predictions owing to high-density radiative-transfer effects \citep[e.g.,][]{Nikopoulos2026, Chang2026}.
According to Table\,2 of \cite{Nikopoulos2026}, JADES-GN-1181\_73488 and CEERS-EGS-1345\_1244 show H$\gamma$/H$\delta$ flux ratios of $\simeq1.4$ and $\simeq1.1$, respectively, comparable to the value inferred from our fit ($1.07^{+0.95}_{-0.47}$).
Moreover, as mentioned in Section\,\ref{sec:spec_fitting}, the wavelength separation between H$\gamma$ and [O\,{\sc iii}]\,$\lambda4363$ is in principle resolvable at the PRISM spectral resolution in their observed wavelength range.

Nevertheless, given the limited $S/N$, we cannot completely exclude the possibility that imperfect deblending of H$\gamma$ and [O\,{\sc iii}]\,$\lambda4363$ substantially affects the inferred line fluxes.
We therefore perform an additional test, assuming the Case B flux ratio between H$\gamma$ and H$\delta$.
If part of the flux attributed to [O\,{\sc iii}]\,$\lambda4363$ in the fiducial fit is instead assigned to H$\gamma$ to satisfy ${\rm H\gamma}/{\rm H\delta}=1.8$, the inferred O3Hg ratio decreases to $-0.21^{+0.43}_{-0.53}$.
This value could also be consistent with non-AGN ionization, depending on the O32 and Ne3O2 ratios, but is comparable to those observed in some LRDs and AGNs \citep{Backhaus2025, Jones2026_4363, Brazzini2026}.
Therefore, the limited wavelength coverage and depth of the available data prevent us from drawing a definitive conclusion about the nature of the ionizing source.

\subsection{Possible iron feature}\label{ssec:iron}
We find that there is also an emission-like feature at $\lambda_{\rm obs}\sim2.7\,\mathchar`-\,2.9\,{\rm \mu m}$, corresponding to $\lambda_{\rm rest}\sim0.24\,\mathchar`-\,0.25\,{\rm \mu m}$ at $z = 10.36$.
Fitting this feature with a single Gaussian (Section\,\ref{sec:spec_fitting}) gives $F_{\rm line}=121.4^{+27.5}_{-27.2}\times10^{-20}\,{\rm erg\,s^{-1}\,cm^{-2}}$ (${\rm EW_{rest}}=79.2^{+18.8}_{-18.7}\,\text{\AA}$, $S/N=4.4$) with a large FWHM of $1.22^{+0.58}_{-0.32}\times10^4\,{\rm km\,s^{-1}}$.
This feature may correspond to the Fe\,{\sc ii} UV bump or the Fe\,{\sc iii} UV47 multiplet, with no other prominent emission lines expected at this wavelength.
Such iron emission lines are typically observed in quasars and are thought to originate from the broad-line region (BLR, e.g., \citealt{Wills1980, Vestergaard2001, Sigut2004}).
The large fitted FWHM is also consistent with an iron-line blend composed of multiple transitions.

Similar iron emission features have been reported in several LRDs \citep[e.g.,][]{Labbe2024, Ji2026_lord, Tripodi2025}, including detections from stacked spectra \citep[e.g.,][]{Perez-Gonzalez2026_stack, Sun2026, Ando2026_UV}.
In contrast, such features have not been commonly reported in normal star-forming galaxies at high redshift. If the observed feature is indeed due to iron emission, it would further support an AGN-powered LRD interpretation for \target\ and may suggest partial UV leakage from the central engine through a clumpy or porous envelope \citep{Ando2026_UV}.
Strong UV iron emission may also provide information on the physical conditions of the nuclear gas, since its strength depends sensitively on gas density, ionization parameter, column density, and radiative-transfer effects \citep[e.g.,][]{Tripodi2025,Torralba2026,Ando2026_UV}. 
If confirmed as iron emission, this feature would suggest substantial chemical enrichment within the first $\sim450\,{\rm Myr}$ of cosmic history.
Both core-collapse supernovae and prompt Type Ia supernovae with short delay times of $\sim30\,\mathchar`-\,100\,{\rm Myr}$ could have contributed to such early enrichment \citep[e.g.,][]{Mannucci2006,Maoz2014,Nakane2024,Nakane2025}.

\subsection{Possible Ly$\alpha$ line}\label{ssec:Lya}
As shown in Figure\,\ref{fig:zoomin}, the IGM+lines model overpredicts the continuum around $\lambda_{\rm obs}=1.4\,\mathchar`-\,1.5\,{\rm \mu m}$, and this remains the case even when fixing $x_{\rm HI}=1.0$, which maximizes the neutral IGM absorption.
When using the IGM+lines model, Ly$\alpha$ emission is not detected ($S/N=0.5$).
In contrast, the IGM+CGM+lines model can reproduce the flux decrement using a strong Ly$\alpha$ damping-wing absorption with $\log\left(N_{\rm HI}/{\rm cm^{-2}}\right) = 22.81^{+0.20}_{-0.25}$ as seen in \cite{Ando2026_UV}.
In this IGM+CGM+lines model, the excess flux around $\lambda_{\rm obs}\sim1.4\,{\rm \mu m}$ can be explained by a Ly$\alpha$ emission line with $S/N=4.2$.
Ly$\alpha$ emission has also been reported in a number of LRDs in previous studies \citep[e.g.,][]{Ji2026_Lya, Kageura2026_Lya, Ando2026_UV}.

A similar spectral feature of a Ly$\alpha$ line with a damping wing that cannot be explained by IGM-only models was reported even in a galaxy at $z = 13.0$ \citep{Wistok2025}.
At such a high redshift, Ly$\alpha$ photons are expected to be strongly attenuated by the substantially neutral IGM; therefore, the detection of a Ly$\alpha$ line requires conditions that enhance Ly$\alpha$ transmission \citep[e.g.,][]{Dijkstra2014_review, Ouchi2020_review}. 
One possibility is that the source resides in a large ionized bubble, within which Ly$\alpha$ photons can redshift away from resonance before encountering the neutral IGM \citep[e.g.,][]{Tang2023_IGM, Umeda2024_IGM}. 
Alternatively, radiative-transfer effects in the ISM and CGM, such as resonant scattering and outflows, can shift Ly$\alpha$ photons redward of the systemic velocity and thereby enhance their transmission even through a smaller ionized region. 

Thus, if the Ly$\alpha$ feature is real, it may indicate a locally ionized environment and/or favorable Ly$\alpha$ radiative-transfer conditions around \target.
However, the significance of the Ly$\alpha$ detection is model-dependent.
Furthermore, the BIC improvement of the IGM+CGM+lines model relative to the IGM+lines model is only $\Delta{\rm BIC}=2.6$, much less significant than the conventional $\Delta{\rm BIC}\gtrsim10$ threshold for strong evidence.
This indicates that the statistical preference for the CGM-included model is weak.
We also note that the downturn around $\lambda_{\rm obs}\sim1.4\,\mathchar`-\,1.5\,{\rm \mu m}$, reproduced by the IGM+CGM model in our fitting, may alternatively be consistent with nebular continuum emission \citep{Ando2026_UV}.
For these reasons, we do not conclude that the CGM-included model is preferred.
Instead, we regard the Ly$\alpha$ detection as model-dependent and refrain from interpreting in detail the fitted parameters sensitive to the Ly$\alpha$ modeling, such as $R_{\rm b}$ and $N_{\rm HI}$.

\section{Discussion: Is it truly an LRD?}\label{sec:discussion}
In Section\,\ref{ssec:spec_z}, we established that the source lies at $z=10.36$ rather than being a low-redshift interloper.
In this section, we discuss whether the observed SED at $z=10.36$ necessarily requires an LRD interpretation, or whether it could instead arise from a non-LRD component.

There is currently no community-wide consensus on a definitive definition of LRDs. 
Nevertheless, three key characteristics are commonly used to identify them: a compact morphology, a V-shaped continuum with a minimum around or redward of the Balmer limit, and broad Balmer emission lines \citep[e.g.,][]{Hviding2025, deGraaff2025_spec_stat}.
For \target, \citet{Tanaka2025_z10} confirmed its compact morphology, and, in this work, we spectroscopically confirm its redshift ($z_{\rm spec}=10.36$, Section\,\ref{ssec:spec_z}) and the presence of the characteristic V-shaped continuum, with its turnover at the Balmer limit (Section\,\ref{ssec:v_shape}).
However, broad Balmer lines have not yet been spectroscopically confirmed because H$\beta$ and H$\alpha$ fall outside the wavelength coverage of the current NIRSpec/PRISM spectrum (Figure\,\ref{fig:spectrum}). 
Although the possibly high [O\,{\sc iii}]\,$\lambda4363$/${\rm H\gamma}$ ratio (Section\,\ref{ssec:line_diag}) and the possible iron emission (Section\,\ref{ssec:iron}) support its LRD or AGN nature, an unambiguous spectroscopic confirmation has not yet been achieved.

We therefore consider whether the observed SED at $z=10.36$ could instead be explained by a stellar-origin Balmer break.
Such a model could in principle reproduce the turnover, and the iron complex and the high O3Hg ratio could arise from an AGN component.
However, this alternative solution is strongly disfavored by the extremely red NIRCam-MIRI color of \target, $m_{\rm F444W}-m_{\rm F770W}=1.82$, based on its robust $S/N=18$ detection in F770W \citep{Tanaka2025_z10}.
If the rest-frame optical continuum were approximately flat in $f_\nu$ from $\lambda_{\rm rest}\sim0.4\,\mathchar`-\,0.9\,{\rm \mu m}$, as expected for a typical Balmer-break galaxy, reproducing the F770W flux would require an H$\alpha$ rest-frame equivalent width (EW) of $\sim(6\pm1)\times10^3\,\text{\AA}$.
Such an extreme EW is substantially larger than those expected theoretically ($\lesssim 4000\,\text{\AA}$; \citealt{Inoue2011}) or typically observed in extreme emission line galaxies ($\sim2000\,\text{\AA}$; e.g., \citealt{Withers2023}). 
The F770W detection therefore provides important support for the LRD nature of \target.
We further confirm that \target\ is independently detected in each of the four individual COSMOS-Web F770W exposures prior to mosaicing, indicating that the F770W signal is unlikely to be caused by an imaging artifact.

Another possible explanation for the red rest-optical continuum is dust-reddened AGN emission \citep[e.g.,][]{Kocevski2023, Matthee2024, Greene2024, Madau2026_dust,Madau2026_polarlization}.
However, the observed turnover near the Balmer limit is characteristic of previously observed LRDs \citep{Setton2024,deGraaff2025_spec_stat}, whereas such a coincidence is not necessarily expected from simple dust reddening.
Furthermore, the tentative C\,{\sc iv} and UV iron emission features, commonly associated with AGN activity, suggest that at least part of the nuclear UV emission is visible without severe obscuration.
This may be in tension with a uniformly dust-obscured AGN, although more complex geometries involving partial covering or scattering, as well as non-standard dust attenuation curves \citep{LiZhengrong2025_dust}, remain possible.
Moreover, despite the ALMA non-detection \citep{Tanaka2025_z10}, obscuration by dust hotter than that effectively probed by the available ALMA observations, or by a compact dust distribution with a small global covering factor, cannot be excluded \citep{Casey2025_dust,Madau2026_dust,Delvecchio2025}.
Similarly, the X-ray non-detection \citep{Tanaka2025_z10} does not rule out the possibility of a Compton-thick AGN.
Therefore, while the observed properties are consistent with an LRD-like interpretation, we cannot rule out dust-obscured AGN scenarios with the current data.
On the other hand, previous studies have not ruled out the possibility that dust attenuation also contributes to the red rest-optical continua of LRDs themselves \citep[e.g.,][]{Madau2026_dust}.
Thus, dust-reddened AGN and other proposed LRD scenarios are not necessarily mutually exclusive, and it remains unclear whether they can be observationally distinguished with the currently available data.
Further observations at longer wavelengths are needed to distinguish between these scenarios and better constrain the physical origin of the red rest-optical continuum.

\section{Conclusion}\label{sec:summary}
\target\ is a robust $z>10$ LRD candidate with $z_{\rm phot}=10.5^{+0.7}_{-0.6}$ selected from the COSMOS-Web dataset using a joint NIRCam-MIRI color selection \citep{Tanaka2025_z10}.
We present the NIRSpec/PRISM spectrum of \target\ (Figure\,\ref{fig:spectrum}) taken in the EMBER program.
The key results are as follows:
\begin{itemize}
    \item Based on the clearly detected Ly$\alpha$ break and emission lines with $S/N\sim2\,\mathchar`-\,4$ (Figure\,\ref{fig:zoomin}), we spectroscopically measure the redshift as $z=10.361^{+0.003}_{-0.002}$, consistent with the photometric redshift (Section\,\ref{ssec:spec_z}).
    The corresponding UV absolute magnitude is $M_{\rm UV}=-20.16^{+0.07}_{-0.07}$, comparable to those of the most UV-luminous LRDs ever found (Figure\,\ref{fig:muv_z}).
    The $z_{\rm spec}$ confirmation removes the principal redshift uncertainty underlying the number density estimates in \cite{Tanaka2025_z10}.
    \item We spectroscopically confirm a V-shaped SED with UV and optical continuum slopes of $\beta_{\rm UV}=-2.26^{+0.11}_{-0.11}$ and $\beta_{\rm opt}=1.17^{+0.74}_{-0.70}$, respectively (Section\,\ref{ssec:v_shape}).
    The turnover wavelength, $\lambda_{\rm turn}=0.356^{+0.010}_{-0.013}\,{\rm \mu m}$, is consistent with the Balmer limit.
    Both of these findings satisfy key spectroscopic characteristics commonly associated with LRDs.
    \item Line detections are tentative with $S/N=2\,\mathchar`-\,4$, and C\,{\sc iv}\,$\lambda\lambda1549,1551$ and [O\,{\sc iii}]\,$\lambda4363$ have the highest significances of $S/N=3.9$ and $3.6$, respectively.
    However, the stacking analysis provides additional support for the line interpretation (Figure\,\ref{fig:stack}, Section\,\ref{ssec:lines}).
    \item The line fitting suggests a possibly high [O\,{\sc iii}]\,$\lambda4363$/H$\gamma$ flux ratio of ${\rm O3Hg} = \log\left({\rm [O\,III]\,\lambda4363}/{\rm H\gamma}\right) = 0.17^{+0.24}_{-0.22}$, or $-0.21^{+0.43}_{-0.53}$ after accounting for possible deblending between [O\,{\sc iii}]\,$\lambda4363$ and H$\gamma$ by imposing the Case B Balmer-line ratio between H$\gamma$ and H$\delta$ (Figure\,\ref{fig:O3Hg_O32}, Section\,\ref{ssec:line_diag}).
    These high line ratios are consistent with those previously observed in LRDs and AGNs, but do not rule out ionization by star formation.
    \item We detect a possible UV iron emission feature at $\lambda_{\rm rest}\sim0.24\,\mathchar`-\,0.25\,{\rm \mu m}$ (Section\,\ref{ssec:iron}).
    If confirmed as iron emission, this feature would be consistent with those observed in previously known AGNs and LRDs. 
    \item We also find a possible, absorption-model-dependent detection of Ly$\alpha$ emission ($S/N=4.2$ with the IGM+CGM model, Section\,\ref{ssec:Lya}) with a strong absorption feature that cannot be reproduced by IGM absorption alone.
    This may suggest enhanced Ly$\alpha$ transmission around \target\ even at $z_{\rm spec}=10.36$.
    \item Due to the lack of spectroscopic confirmation of broad Balmer lines, the LRD nature of \target\ has not yet been established definitively. 
    Nevertheless, when combined with the previously reported large $m_{\rm F444W}-m_{\rm F770W}$ color excess and the spectroscopic features presented here, the LRD interpretation provides the most plausible explanation for the observed properties of \target\ (Section\,\ref{sec:discussion}).
    We therefore regard \target\ as the first $z>10$ LRD candidate with a spectroscopically confirmed redshift. 
\end{itemize}

The planned MIRI/LRS spectroscopy in Cycle 5 (GO\#9483, PI T.~S.~Tanaka) will provide a deep mid-infrared spectrum covering H$\alpha$, H$\beta$, [O\,{\sc iii}]\,$\lambda5007$ and the rest-optical continuum, which will enable further tests and a more detailed characterization of its nature.
Complementing the MIRI spectroscopy, deeper NIRSpec observations at higher spectral resolution will enable more detailed spectroscopic characterization through more robust line detections, improved Ly$\alpha$ and IGM+CGM modeling, and more reliable photoionization modeling.

To constrain the evolution of the LRD population, we need a larger $z>10$ LRD sample, which requires more extensive MIRI observations.
Beyond \JWST-only searches, future MIRI follow-up of bright $z>10$ candidates identified by the recently launched \textit{Roman} Space Telescope will be effective for searching for more luminous $z>10$ LRDs over much wider areas.
On a longer timescale, the proposed \textit{GREX-PLUS} mission will further extend such searches through its wide-field near- to mid-infrared imaging capabilities \citep{GREXPLUS_SB,GREXPLUS_SPIE2024,GREX_SB_v2}.

\begin{acknowledgements}
This work is based on observations made with the NASA/ESA/CSA \textit{James Webb Space Telescope}. The data were obtained from the Mikulski Archive for Space Telescopes at the Space Telescope Science Institute, which is operated by the Association of Universities for Research in Astronomy, Inc., under NASA contract NAS5-03127 for JWST. These observations are associated with program GO-7076.
Support for US investigators in program GO-7076 was provided by NASA through a grant from the Space Telescope Science Institute, which is operated by the Association of Universities for Research in Astronomy, Inc., under NASA contract NAS5-03127.
The observation can be accessed via \dataset[doi: 10.17909/7mg3-kh29]{\doi{10.17909/7mg3-kh29}}.
Numerical computations were in part carried out on the iDark cluster, Kavli IPMU.
This work was made possible by utilizing the CANDIDE cluster at the Institut d’Astrophysique de Paris.
The CANDIDE cluster was funded through grants from the PNCG, CNES, DIM-ACAV, the Euclid Consortium, and the Danish National Research Foundation Cosmic Dawn Center (DNRF140) and is maintained by Stephane Rouberol.
The COSMOS team is partly supported by the Centre National d’Etudes Spatiales (CNES).
Kavli IPMU is supported by World Premier International Research Center Initiative (WPI), MEXT, Japan.
The Cosmic Dawn Center is funded by the Danish National Research Foundation (DNRF140).
We thank Yoshihisa~Asada, Ignas~Juodžbalis, and Lucy~R.~Ivey for fruitful discussion.
We thank the anonymous referee for helpful feedback.
T.~S.~Tanaka is supported by Japan Society for the Promotion of Science (JSPS) KAKENHI Grant Number JP25KJ0750 and the Forefront Physics and Mathematics Program to Drive Transformation (FoPM), a World-leading Innovative Graduate Study (WINGS) Program at the University of Tokyo.
T.~S.~Tanaka gratefully acknowledges UNISON SQUARE GARDEN, whose full-color music motivated him throughout this research and inspired him to keep surprising the world.
M.~Aravena is supported by FONDECYT grant number 1252054, and gratefully acknowledges support from ANID Basal Project FB210003,  ANID MILENIO NCN2024\_112 and ANID + Vinculaci\'on Internacional + FOVI250261.
B.~Trakhtenbrot acknowledges support from the European Research Council (ERC) under the European Union’s Horizon 2020 research and innovation program (grant agreement No. 950533).
K.~Inayoshi acknowledges support from the National Natural Science Foundation of China (12573015, W2532003), the Beijing Natural Science Foundation (IS25003), and the China Manned Space Program (CMS-CSST-2025-A09).
J.-T.~Schindler acknowledges funding by the Deutsche Forschungsgemeinschaft (DFG, German Research Foundation) - Project number 518006966 - and support by the DFG under Germany’s Excellence Strategy – EXC 2121 „Quantum Universe“ – 390833306.
A.~K.~Inoue is supported by Japan Society for the Promotion of Science (JSPS) KAKENHI Grant Number JP26H02069.
O.~Ilbert acknowledges the funding of the French Agence Nationale de la Recherche for the project iMAGE (grant ANR-22-CE31-0007).
D.~B.~Sanders gratefully acknowledges support from NSF Grant 2407752.
E.~Treister acknowledges support from the ANID CATA-BASAL program FB210003, and FONDECYT Regular 1241005 and  1250821.
OpenAI ChatGPT was used under author supervision to assist with language editing and refinement of the manuscript text. 
All scientific content, interpretations, references, and final wording were reviewed and approved by the authors, who take full responsibility for the content of the manuscript.
\end{acknowledgements}

\newpage
\bibliographystyle{aasjournalv7} 
\bibliography{z10LRD}

@ARTICLE{Delvecchio2025,
       author = {{Delvecchio}, I. and {Daddi}, E. and {Magnelli}, B. and {Elbaz}, D. and {Giavalisco}, M. and {Traina}, A. and {Lanzuisi}, G. and {Akins}, H.~B. and {Belli}, S. and {Casey}, C.~M. and {Gentile}, F. and {Gruppioni}, C. and {Pozzi}, F. and {Zamorani}, G.},
        title = "{Active galactic nuclei-heated dust revealed in ``little red dots''}",
      journal = {\aap},
         year = 2025,
        month = dec,
       volume = {704},
          eid = {A313},
        pages = {A313},
          doi = {10.1051/0004-6361/202557164},
archivePrefix = {arXiv},
       eprint = {2509.07100},
 primaryClass = {astro-ph.GA},
       adsurl = {https://ui.adsabs.harvard.edu/abs/2025A&A...704A.313D}
}

@ARTICLE{Lambrides2026_variability,
       author = {{Lambrides}, Erini and {Hutchison}, Taylor A. and {Larson}, Rebecca L. and {Arrabal Haro}, Pablo and {Papovich}, Casey and {Hu}, Weida and {Cleri}, Nikko J. and {Finkelstein}, Steven L. and {Trump}, Jonathan R. and {Perez-Gonzalez}, Pablo G. and {Wang}, Bingjie and {Kocevski}, Dale D. and {Chisholm}, John and {Secunda}, Amy and {Bosman}, Sarah E.~I. and {Akins}, Hollis and {Karmen}, Mitchell and {Dickinson}, Mark and {Bromm}, Volker and {Backhaus}, Bren E. and {Chiaberge}, Marco and {Cooper}, Olivia R. and {Ajay}, Yukta and {Barro}, Guillermo and {Berg}, Danielle A. and {Cann}, Jenna and {Cooper}, M.~C. and {Grogin}, Norman A. and {Hirschmann}, Michaela and {Huertas-Company}, Marc and {Kartaltepe}, Jeyhan S. and {Koekemoer}, Anton M. and {Lucas}, Ray A. and {Long}, Arianna S. and {Gilli}, Roberto and {Norman}, Colin and {Ptak}, Andrew F. and {Richardson}, Chris T. and {Rigby}, Jane R. and {Vanderhoof}, Brittany N. and {Yung}, L.~Y. Aaron and {Zavala}, Jorge A.},
        title = "{The GlimmIr: Spectroscopic Variability in a z\raisebox{-0.5ex}\textasciitilde7 LRD Indicates Rapid Changes in Both the Narrow and Broad Line Regions}",
      journal = {arXiv e-prints},
         year = 2026,
        month = apr,
          eid = {arXiv:2604.25991},
        pages = {arXiv:2604.25991},
          doi = {10.48550/arXiv.2604.25991},
archivePrefix = {arXiv},
       eprint = {2604.25991},
 primaryClass = {astro-ph.GA},
       adsurl = {https://ui.adsabs.harvard.edu/abs/2026arXiv260425991L}
}

@ARTICLE{Brooks2026_variability,
       author = {{Brooks}, Madisyn and {Davis}, Kelcey and {Trump}, Jonathan R. and {Simons}, Raymond C. and {Lambrides}, Erini and {Arrabal Haro}, Pablo and {Backhaus}, Bren E. and {Cleri}, Nikko J. and {Finkelstein}, Steven L. and {Giavalisco}, Mauro and {Grogin}, Norman A. and {Hirschmann}, Michaela and {Kocevski}, Dale D. and {Koekemoer}, Anton M. and {Larson}, Rebecca L. and {Lucas}, Ray A. and {Wilkins}, Stephen M. and {Wuyts}, Stijn and {Zavala}, Jorge A.},
        title = "{Ripples in the OCEANS: Broad Line Variability of Little Red Dots}",
      journal = {arXiv e-prints},
         year = 2026,
        month = aug,
          eid = {arXiv:2608.12487},
        pages = {arXiv:2608.12487},
          doi = {10.48550/arXiv.2608.12487},
archivePrefix = {arXiv},
       eprint = {2608.12487},
 primaryClass = {astro-ph.GA},
       adsurl = {https://ui.adsabs.harvard.edu/abs/2026arXiv260812487B}
}

@ARTICLE{Johnson2021,
       author = {{Johnson}, Benjamin D. and {Leja}, Joel and {Conroy}, Charlie and {Speagle}, Joshua S.},
        title = "{Stellar Population Inference with Prospector}",
      journal = {\apjs},
         year = 2021,
        month = jun,
       volume = {254},
       number = {2},
          eid = {22},
        pages = {22},
          doi = {10.3847/1538-4365/abef67},
archivePrefix = {arXiv},
       eprint = {2012.01426},
 primaryClass = {astro-ph.GA},
       adsurl = {https://ui.adsabs.harvard.edu/abs/2021ApJS..254...22J}
}

@ARTICLE{Kormendy2013,
       author = {{Kormendy}, John and {Ho}, Luis C.},
        title = "{Coevolution (Or Not) of Supermassive Black Holes and Host Galaxies}",
      journal = {\araa},
         year = 2013,
        month = aug,
       volume = {51},
       number = {1},
        pages = {511-653},
          doi = {10.1146/annurev-astro-082708-101811},
archivePrefix = {arXiv},
       eprint = {1304.7762},
 primaryClass = {astro-ph.CO},
       adsurl = {https://ui.adsabs.harvard.edu/abs/2013ARA&A..51..511K}
}

@ARTICLE{Lin2025_lowz,
       author = {{Lin}, Xiaojing and {Fan}, Xiaohui and {Cai}, Zheng and {Bian}, Fuyan and {Liu}, Hanpu and {Sun}, Fengwu and {Ma}, Yilun and {Greene}, Jenny E. and {Strauss}, Michael A. and {Green}, Richard and {Lyu}, Jianwei and {Champagne}, Jaclyn B. and {Goulding}, Andy D. and {Inayoshi}, Kohei and {Jin}, Xiangyu and {Leung}, Gene C.~K. and {Li}, Mingyu and {Liu}, Weizhe and {Liu}, Yichen and {Mao}, Junjie and {Pudoka}, Maria Anne and {Tee}, Wei Leong and {Wang}, Ben and {Wang}, Feige and {Wu}, Yunjing and {Yang}, Jinyi and {Zhang}, Haowen and {Zhu}, Yongda},
        title = "{The Discovery of Little Red Dots in the Local Universe: Signatures of Cool Gas Envelopes}",
      journal = {\apj},
         year = 2026,
        month = feb,
       volume = {997},
       number = {2},
          eid = {364},
        pages = {364},
          doi = {10.3847/1538-4357/ae2bdf},
archivePrefix = {arXiv},
       eprint = {2507.10659},
 primaryClass = {astro-ph.GA},
       adsurl = {https://ui.adsabs.harvard.edu/abs/2026ApJ...997..364L}
}

@ARTICLE{Casey2023,
       author = {{Casey}, Caitlin M. and {Kartaltepe}, Jeyhan S. and {Drakos}, Nicole E. and {Franco}, Maximilien and {Harish}, Santosh and {Paquereau}, Louise and {Ilbert}, Olivier and {Rose}, Caitlin and {Cox}, Isabella G. and {Nightingale}, James W. and {Robertson}, Brant E. and {Silverman}, John D. and {Koekemoer}, Anton M. and {Massey}, Richard and {McCracken}, Henry Joy and {Rhodes}, Jason and {Akins}, Hollis B. and {Allen}, Natalie and {Amvrosiadis}, Aristeidis and {Arango-Toro}, Rafael C. and {Bagley}, Micaela B. and {Bongiorno}, Angela and {Capak}, Peter L. and {Champagne}, Jaclyn B. and {Chartab}, Nima and {Ch{\'a}vez Ortiz}, {\'O}scar A. and {Chworowsky}, Katherine and {Cooke}, Kevin C. and {Cooper}, Olivia R. and {Darvish}, Behnam and {Ding}, Xuheng and {Faisst}, Andreas L. and {Finkelstein}, Steven L. and {Fujimoto}, Seiji and {Gentile}, Fabrizio and {Gillman}, Steven and {Gould}, Katriona M.~L. and {Gozaliasl}, Ghassem and {Hayward}, Christopher C. and {He}, Qiuhan and {Hemmati}, Shoubaneh and {Hirschmann}, Michaela and {Jahnke}, Knud and {Jin}, Shuowen and {Khostovan}, Ali Ahmad and {Kokorev}, Vasily and {Lambrides}, Erini and {Laigle}, Clotilde and {Larson}, Rebecca L. and {Leung}, Gene C.~K. and {Liu}, Daizhong and {Liaudat}, Tobias and {Long}, Arianna S. and {Magdis}, Georgios and {Mahler}, Guillaume and {Mainieri}, Vincenzo and {Manning}, Sinclaire M. and {Maraston}, Claudia and {Martin}, Crystal L. and {McCleary}, Jacqueline E. and {McKinney}, Jed and {McPartland}, Conor J.~R. and {Mobasher}, Bahram and {Pattnaik}, Rohan and {Renzini}, Alvio and {Rich}, R. Michael and {Sanders}, David B. and {Sattari}, Zahra and {Scognamiglio}, Diana and {Scoville}, Nick and {Sheth}, Kartik and {Shuntov}, Marko and {Sparre}, Martin and {Suzuki}, Tomoko L. and {Talia}, Margherita and {Toft}, Sune and {Trakhtenbrot}, Benny and {Urry}, C. Megan and {Valentino}, Francesco and {Vanderhoof}, Brittany N. and {Vardoulaki}, Eleni and {Weaver}, John R. and {Whitaker}, Katherine E. and {Wilkins}, Stephen M. and {Yang}, Lilan and {Zavala}, Jorge A.},
        title = "{COSMOS-Web: An Overview of the JWST Cosmic Origins Survey}",
      journal = {\apj},
         year = 2023,
        month = sep,
       volume = {954},
       number = {1},
          eid = {31},
        pages = {31},
          doi = {10.3847/1538-4357/acc2bc},
archivePrefix = {arXiv},
       eprint = {2211.07865},
 primaryClass = {astro-ph.GA},
       adsurl = {https://ui.adsabs.harvard.edu/abs/2023ApJ...954...31C}
}

@ARTICLE{Matthee2024,
       author = {{Matthee}, Jorryt and {Naidu}, Rohan P. and {Brammer}, Gabriel and {Chisholm}, John and {Eilers}, Anna-Christina and {Goulding}, Andy and {Greene}, Jenny and {Kashino}, Daichi and {Labbe}, Ivo and {Lilly}, Simon J. and {Mackenzie}, Ruari and {Oesch}, Pascal A. and {Weibel}, Andrea and {Wuyts}, Stijn and {Xiao}, Mengyuan and {Bordoloi}, Rongmon and {Bouwens}, Rychard and {van Dokkum}, Pieter and {Illingworth}, Garth and {Kramarenko}, Ivan and {Maseda}, Michael V. and {Mason}, Charlotte and {Meyer}, Romain A. and {Nelson}, Erica J. and {Reddy}, Naveen A. and {Shivaei}, Irene and {Simcoe}, Robert A. and {Yue}, Minghao},
        title = "{Little Red Dots: An Abundant Population of Faint Active Galactic Nuclei at z {\ensuremath{\sim}} 5 Revealed by the EIGER and FRESCO JWST Surveys}",
      journal = {\apj},
         year = 2024,
        month = mar,
       volume = {963},
       number = {2},
          eid = {129},
        pages = {129},
          doi = {10.3847/1538-4357/ad2345},
archivePrefix = {arXiv},
       eprint = {2306.05448},
 primaryClass = {astro-ph.GA},
       adsurl = {https://ui.adsabs.harvard.edu/abs/2024ApJ...963..129M}
}

@ARTICLE{Kocevski2026,
       author = {{Kocevski}, Dale D. and {Taylor}, Anthony J. and {Onoue}, Masafusa and {Inayoshi}, Kohei and {Finkelstein}, Steven L. and {Barro}, Guillermo and {Guo}, Jingsong and {Arrabal Haro}, Pablo and {Trump}, Jonathan R. and {Larson}, Rebecca L. and {Dickinson}, Mark and {Papovich}, Casey and {Pacucci}, Fabio and {Perez-Gonzalez}, Pablo G. and {Hirschmann}, Michaela and {McGrath}, Elizabeth J. and {Jones}, Brenda L. and {Cleri}, Nikko J. and {Davis}, Kelcey and {Giavalisco}, Mauro and {Grogin}, Norman A. and {Hutchison}, Taylor A. and {Kartaltepe}, Jeyhan S. and {Kirkpatrick}, Allison and {Leung}, Gene C.~K. and {Somerville}, Rachel S. and {Wilkins}, Stephen M.},
        title = "{Skyfire: A Spectroscopic Census of Little Red Dots and Broad-Line AGN in the CEERS Field}",
      journal = {arXiv e-prints},
         year = 2026,
        month = aug,
          eid = {arXiv:2609.00112},
        pages = {arXiv:2609.00112},
archivePrefix = {arXiv},
       eprint = {2609.00112},
 primaryClass = {astro-ph.GA},
       adsurl = {https://ui.adsabs.harvard.edu/abs/2026arXiv260900112K}
}

@ARTICLE{Barro2024,
       author = {{Barro}, Guillermo and {P{\'e}rez-Gonz{\'a}lez}, Pablo G. and {Kocevski}, Dale D. and {McGrath}, Elizabeth J. and {Trump}, Jonathan R. and {Simons}, Raymond C. and {Somerville}, Rachel S. and {Yung}, L.~Y. Aaron and {Arrabal Haro}, Pablo and {Akins}, Hollis B. and {Bagley}, Michaela B. and {Cleri}, Nikko J. and {Costantin}, Luca and {Davis}, Kelcey and {Dickinson}, Mark and {Finkelstein}, Steve L. and {Giavalisco}, Mauro and {G{\'o}mez-Guijarro}, Carlos and {Hathi}, Nimish P. and {Hirschmann}, Michaela and {Holwerda}, Benne W. and {Huertas-Company}, Marc and {Kartaltepe}, Jeyhan S. and {Koekemoer}, Anton M. and {Lucas}, Ray A. and {Papovich}, Casey and {Pirzkal}, Nor and {Seill{\'e}}, Lise-Marie and {Tacchella}, Sandro and {Wuyts}, Stijn and {Wilkins}, Stephen M. and {de la Vega}, Alexander and {Yang}, Guang and {Zavala}, Jorge A.},
        title = "{Extremely Red Galaxies at z = 5{\textendash}9 with MIRI and NIRSpec: Dusty Galaxies or Obscured Active Galactic Nuclei?}",
      journal = {\apj},
         year = 2024,
        month = mar,
       volume = {963},
       number = {2},
          eid = {128},
        pages = {128},
          doi = {10.3847/1538-4357/ad167e},
archivePrefix = {arXiv},
       eprint = {2305.14418},
 primaryClass = {astro-ph.GA},
       adsurl = {https://ui.adsabs.harvard.edu/abs/2024ApJ...963..128B}
}

@ARTICLE{Labbe2023b,
       author = {{Labbe}, Ivo and {Greene}, Jenny E. and {Bezanson}, Rachel and {Fujimoto}, Seiji and {Furtak}, Lukas J. and {Goulding}, Andy D. and {Matthee}, Jorryt and {Naidu}, Rohan P. and {Oesch}, Pascal A. and {Atek}, Hakim and {Brammer}, Gabriel and {Chemerynska}, Iryna and {Coe}, Dan and {Cutler}, Sam E. and {Dayal}, Pratika and {Feldmann}, Robert and {Franx}, Marijn and {Glazebrook}, Karl and {Leja}, Joel and {Maseda}, Michael and {Marchesini}, Danilo and {Nanayakkara}, Themiya and {Nelson}, Erica J. and {Pan}, Richard and {Papovich}, Casey and {Price}, Sedona H. and {Suess}, Katherine A. and {Wang}, Bingjie and {Weaver}, John R. and {Whitaker}, Katherine E. and {Williams}, Christina C. and {Zitrin}, Adi},
        title = "{UNCOVER: Candidate Red Active Galactic Nuclei at 3 < z < 7 with JWST and ALMA}",
      journal = {\apj},
         year = 2025,
        month = jan,
       volume = {978},
       number = {1},
          eid = {92},
        pages = {92},
          doi = {10.3847/1538-4357/ad3551},
archivePrefix = {arXiv},
       eprint = {2306.07320},
 primaryClass = {astro-ph.GA},
       adsurl = {https://ui.adsabs.harvard.edu/abs/2025ApJ...978...92L}
}

@ARTICLE{Kocevski2023,
       author = {{Kocevski}, Dale D. and {Onoue}, Masafusa and {Inayoshi}, Kohei and {Trump}, Jonathan R. and {Arrabal Haro}, Pablo and {Grazian}, Andrea and {Dickinson}, Mark and {Finkelstein}, Steven L. and {Kartaltepe}, Jeyhan S. and {Hirschmann}, Michaela and {Aird}, James and {Holwerda}, Benne W. and {Fujimoto}, Seiji and {Juneau}, St{\'e}phanie and {Amor{\'\i}n}, Ricardo O. and {Backhaus}, Bren E. and {Bagley}, Micaela B. and {Barro}, Guillermo and {Bell}, Eric F. and {Bisigello}, Laura and {Calabr{\`o}}, Antonello and {Cleri}, Nikko J. and {Cooper}, M.~C. and {Ding}, Xuheng and {Grogin}, Norman A. and {Ho}, Luis C. and {Hutchison}, Taylor A. and {Inoue}, Akio K. and {Jiang}, Linhua and {Jones}, Brenda and {Koekemoer}, Anton M. and {Li}, Wenxiu and {Li}, Zhengrong and {McGrath}, Elizabeth J. and {Molina}, Juan and {Papovich}, Casey and {P{\'e}rez-Gonz{\'a}lez}, Pablo G. and {Pirzkal}, Nor and {Wilkins}, Stephen M. and {Yang}, Guang and {Yung}, L.~Y. Aaron},
        title = "{Hidden Little Monsters: Spectroscopic Identification of Low-mass, Broad-line AGNs at z > 5 with CEERS}",
      journal = {\apjl},
         year = 2023,
        month = sep,
       volume = {954},
       number = {1},
          eid = {L4},
        pages = {L4},
          doi = {10.3847/2041-8213/ace5a0},
archivePrefix = {arXiv},
       eprint = {2302.00012},
 primaryClass = {astro-ph.GA},
       adsurl = {https://ui.adsabs.harvard.edu/abs/2023ApJ...954L...4K}
}

@ARTICLE{Kocevski2024,
       author = {{Kocevski}, Dale D. and {Finkelstein}, Steven L. and {Barro}, Guillermo and {Taylor}, Anthony J. and {Calabr{\`o}}, Antonello and {Laloux}, Brivael and {Buchner}, Johannes and {Trump}, Jonathan R. and {Leung}, Gene C.~K. and {Yang}, Guang and {Dickinson}, Mark and {P{\'e}rez-Gonz{\'a}lez}, Pablo G. and {Pacucci}, Fabio and {Inayoshi}, Kohei and {Somerville}, Rachel S. and {McGrath}, Elizabeth J. and {Akins}, Hollis B. and {Bagley}, Micaela B. and {Bowler}, Rebecca A.~A. and {Bisigello}, Laura and {Carnall}, Adam and {Casey}, Caitlin M. and {Cheng}, Yingjie and {Cleri}, Nikko J. and {Costantin}, Luca and {Cullen}, Fergus and {Davis}, Kelcey and {Donnan}, Callum T. and {Dunlop}, James S. and {Ellis}, Richard S. and {Ferguson}, Henry C. and {Fujimoto}, Seiji and {Fontana}, Adriano and {Giavalisco}, Mauro and {Grazian}, Andrea and {Grogin}, Norman A. and {Hathi}, Nimish P. and {Hirschmann}, Michaela and {Huertas-Company}, Marc and {Holwerda}, Benne W. and {Illingworth}, Garth and {Juneau}, St{\'e}phanie and {Kartaltepe}, Jeyhan S. and {Koekemoer}, Anton M. and {Li}, Wenxiu and {Lucas}, Ray A. and {Magee}, Dan and {Mason}, Charlotte and {McLeod}, Derek J. and {McLure}, Ross J. and {Napolitano}, Lorenzo and {Papovich}, Casey and {Pirzkal}, Nor and {Rodighiero}, Giulia and {Santini}, Paola and {Wilkins}, Stephen M. and {Yung}, L.~Y. Aaron},
        title = "{The Rise of Faint, Red Active Galactic Nuclei at z > 4: A Sample of Little Red Dots in the JWST Extragalactic Legacy Fields}",
      journal = {\apj},
         year = 2025,
        month = jun,
       volume = {986},
       number = {2},
          eid = {126},
        pages = {126},
          doi = {10.3847/1538-4357/adbc7d},
archivePrefix = {arXiv},
       eprint = {2404.03576},
 primaryClass = {astro-ph.GA},
       adsurl = {https://ui.adsabs.harvard.edu/abs/2025ApJ...986..126K}
}

@ARTICLE{Greene2024,
       author = {{Greene}, Jenny E. and {Labbe}, Ivo and {Goulding}, Andy D. and {Furtak}, Lukas J. and {Chemerynska}, Iryna and {Kokorev}, Vasily and {Dayal}, Pratika and {Volonteri}, Marta and {Williams}, Christina C. and {Wang}, Bingjie and {Setton}, David J. and {Burgasser}, Adam J. and {Bezanson}, Rachel and {Atek}, Hakim and {Brammer}, Gabriel and {Cutler}, Sam E. and {Feldmann}, Robert and {Fujimoto}, Seiji and {Glazebrook}, Karl and {de Graaff}, Anna and {Khullar}, Gourav and {Leja}, Joel and {Marchesini}, Danilo and {Maseda}, Michael V. and {Matthee}, Jorryt and {Miller}, Tim B. and {Naidu}, Rohan P. and {Nanayakkara}, Themiya and {Oesch}, Pascal A. and {Pan}, Richard and {Papovich}, Casey and {Price}, Sedona H. and {van Dokkum}, Pieter and {Weaver}, John R. and {Whitaker}, Katherine E. and {Zitrin}, Adi},
        title = "{UNCOVER Spectroscopy Confirms the Surprising Ubiquity of Active Galactic Nuclei in Red Sources at z > 5}",
      journal = {\apj},
         year = 2024,
        month = mar,
       volume = {964},
       number = {1},
          eid = {39},
        pages = {39},
          doi = {10.3847/1538-4357/ad1e5f},
archivePrefix = {arXiv},
       eprint = {2309.05714},
 primaryClass = {astro-ph.GA},
       adsurl = {https://ui.adsabs.harvard.edu/abs/2024ApJ...964...39G}
}

@ARTICLE{Furtak2024,
       author = {{Furtak}, Lukas J. and {Labb{\'e}}, Ivo and {Zitrin}, Adi and {Greene}, Jenny E. and {Dayal}, Pratika and {Chemerynska}, Iryna and {Kokorev}, Vasily and {Miller}, Tim B. and {Goulding}, Andy D. and {de Graaff}, Anna and {Bezanson}, Rachel and {Brammer}, Gabriel B. and {Cutler}, Sam E. and {Leja}, Joel and {Pan}, Richard and {Price}, Sedona H. and {Wang}, Bingjie and {Weaver}, John R. and {Whitaker}, Katherine E. and {Atek}, Hakim and {Bogd{\'a}n}, {\'A}kos and {Charlot}, St{\'e}phane and {Curtis-Lake}, Emma and {van Dokkum}, Pieter and {Endsley}, Ryan and {Feldmann}, Robert and {Fudamoto}, Yoshinobu and {Fujimoto}, Seiji and {Glazebrook}, Karl and {Juneau}, St{\'e}phanie and {Marchesini}, Danilo and {Maseda}, Micheal V. and {Nelson}, Erica and {Oesch}, Pascal A. and {Plat}, Ad{\`e}le and {Setton}, David J. and {Stark}, Daniel P. and {Williams}, Christina C.},
        title = "{A high black-hole-to-host mass ratio in a lensed AGN in the early Universe}",
      journal = {\nat},
         year = 2024,
        month = apr,
       volume = {628},
       number = {8006},
        pages = {57},
          doi = {10.1038/s41586-024-07184-8},
       adsurl = {https://ui.adsabs.harvard.edu/abs/2024Natur.628...57F}
}

@ARTICLE{Harikane2023_agn,
       author = {{Harikane}, Yuichi and {Zhang}, Yechi and {Nakajima}, Kimihiko and {Ouchi}, Masami and {Isobe}, Yuki and {Ono}, Yoshiaki and {Hatano}, Shun and {Xu}, Yi and {Umeda}, Hiroya},
        title = "{A JWST/NIRSpec First Census of Broad-line AGNs at z = 4-7: Detection of 10 Faint AGNs with M $_{BH}$ {}10$^{6}$-{}10$^{8}$ M $_{{\ensuremath{\odot}}}$ and Their Host Galaxy Properties}",
      journal = {\apj},
         year = 2023,
        month = dec,
       volume = {959},
       number = {1},
          eid = {39},
        pages = {39},
          doi = {10.3847/1538-4357/ad029e},
archivePrefix = {arXiv},
       eprint = {2303.11946},
 primaryClass = {astro-ph.GA},
       adsurl = {https://ui.adsabs.harvard.edu/abs/2023ApJ...959...39H}
}

@ARTICLE{Yue2024,
       author = {{Yue}, Minghao and {Eilers}, Anna-Christina and {Ananna}, Tonima Tasnim and {Panagiotou}, Christos and {Kara}, Erin and {Miyaji}, Takamitsu},
        title = "{Stacking X-Ray Observations of ``Little Red Dots'': Implications for Their Active Galactic Nucleus Properties}",
      journal = {\apjl},
         year = 2024,
        month = oct,
       volume = {974},
       number = {2},
          eid = {L26},
        pages = {L26},
          doi = {10.3847/2041-8213/ad7eba},
archivePrefix = {arXiv},
       eprint = {2404.13290},
 primaryClass = {astro-ph.GA},
       adsurl = {https://ui.adsabs.harvard.edu/abs/2024ApJ...974L..26Y}
}

@ARTICLE{Ananna2024,
       author = {{Ananna}, Tonima Tasnim and {Bogd{\'a}n}, {\'A}kos and {Kov{\'a}cs}, Orsolya E. and {Natarajan}, Priyamvada and {Hickox}, Ryan C.},
        title = "{X-Ray View of Little Red Dots: Do They Host Supermassive Black Holes?}",
      journal = {\apjl},
         year = 2024,
        month = jul,
       volume = {969},
       number = {1},
          eid = {L18},
        pages = {L18},
          doi = {10.3847/2041-8213/ad5669},
archivePrefix = {arXiv},
       eprint = {2404.19010},
 primaryClass = {astro-ph.GA},
       adsurl = {https://ui.adsabs.harvard.edu/abs/2024ApJ...969L..18A}
}

@ARTICLE{Labbe2023a,
       author = {{Labb{\'e}}, Ivo and {van Dokkum}, Pieter and {Nelson}, Erica and {Bezanson}, Rachel and {Suess}, Katherine A. and {Leja}, Joel and {Brammer}, Gabriel and {Whitaker}, Katherine and {Mathews}, Elijah and {Stefanon}, Mauro and {Wang}, Bingjie},
        title = "{A population of red candidate massive galaxies  600 Myr after the Big Bang}",
      journal = {\nat},
         year = 2023,
        month = apr,
       volume = {616},
       number = {7956},
        pages = {266-269},
          doi = {10.1038/s41586-023-05786-2},
archivePrefix = {arXiv},
       eprint = {2207.12446},
 primaryClass = {astro-ph.GA},
       adsurl = {https://ui.adsabs.harvard.edu/abs/2023Natur.616..266L}
}

@ARTICLE{Maiolino2024_chandra,
       author = {{Maiolino}, Roberto and {Risaliti}, Guido and {Signorini}, Matilde and {Trefoloni}, Bartolomeo and {Juod{\v{z}}balis}, Ignas and {Scholtz}, Jan and {{\"U}bler}, Hannah and {D'Eugenio}, Francesco and {Carniani}, Stefano and {Fabian}, Andy and {Ji}, Xihan and {Mazzolari}, Giovanni and {Bertola}, Elena and {Brusa}, Marcella and {Bunker}, Andrew J. and {Charlot}, Stephane and {Comastri}, Andrea and {Cresci}, Giovanni and {DeCoursey}, Christa Noel and {Egami}, Eiichi and {Fiore}, Fabrizio and {Gilli}, Roberto and {Perna}, Michele and {Tacchella}, Sandro and {Venturi}, Giacomo},
        title = "{JWST meets Chandra: a large population of Compton thick, feedback-free, and intrinsically X-ray weak AGN, with a sprinkle of SNe}",
      journal = {\mnras},
         year = 2025,
        month = apr,
       volume = {538},
       number = {3},
        pages = {1921-1943},
          doi = {10.1093/mnras/staf359},
archivePrefix = {arXiv},
       eprint = {2405.00504},
 primaryClass = {astro-ph.GA},
       adsurl = {https://ui.adsabs.harvard.edu/abs/2025MNRAS.538.1921M}
}

@ARTICLE{Akins2024,
       author = {{Akins}, Hollis B. and {Casey}, Caitlin M. and {Lambrides}, Erini and {Allen}, Natalie and {Andika}, Irham T. and {Brinch}, Malte and {Champagne}, Jaclyn B. and {Cooper}, Olivia and {Ding}, Xuheng and {Drakos}, Nicole E. and {Faisst}, Andreas and {Finkelstein}, Steven L. and {Franco}, Maximilien and {Fujimoto}, Seiji and {Gentile}, Fabrizio and {Gillman}, Steven and {Gozaliasl}, Ghassem and {Harish}, Santosh and {Hayward}, Christopher C. and {Hirschmann}, Michaela and {Ilbert}, Olivier and {Kartaltepe}, Jeyhan S. and {Kocevski}, Dale D. and {Koekemoer}, Anton M. and {Kokorev}, Vasily and {Liu}, Daizhong and {Long}, Arianna S. and {McCracken}, Henry Joy and {McKinney}, Jed and {Onoue}, Masafusa and {Paquereau}, Louise and {Renzini}, Alvio and {Rhodes}, Jason and {Robertson}, Brant E. and {Shuntov}, Marko and {Silverman}, John D. and {Tanaka}, Takumi S. and {Toft}, Sune and {Trakhtenbrot}, Benny and {Valentino}, Francesco and {Zavala}, Jorge},
        title = "{COSMOS-Web: The Overabundance and Physical Nature of ``Little Red Dots''{\textemdash}Implications for Early Galaxy and SMBH Assembly}",
      journal = {\apj},
         year = 2025,
        month = sep,
       volume = {991},
       number = {1},
          eid = {37},
        pages = {37},
          doi = {10.3847/1538-4357/ade984},
archivePrefix = {arXiv},
       eprint = {2406.10341},
 primaryClass = {astro-ph.GA},
       adsurl = {https://ui.adsabs.harvard.edu/abs/2025ApJ...991...37A}
}

@ARTICLE{InayoshiMaiolino2024,
       author = {{Inayoshi}, Kohei and {Maiolino}, Roberto},
        title = "{Extremely Dense Gas around Little Red Dots and High-redshift Active Galactic Nuclei: A Nonstellar Origin of the Balmer Break and Absorption Features}",
      journal = {\apjl},
         year = 2025,
        month = feb,
       volume = {980},
       number = {2},
          eid = {L27},
        pages = {L27},
          doi = {10.3847/2041-8213/adaebd},
archivePrefix = {arXiv},
       eprint = {2409.07805},
 primaryClass = {astro-ph.GA},
       adsurl = {https://ui.adsabs.harvard.edu/abs/2025ApJ...980L..27I}
}

@ARTICLE{Kokubo2024,
       author = {{Kokubo}, Mitsuru and {Harikane}, Yuichi},
        title = "{Challenging the Active Galactic Nucleus Scenario for JWST/NIRSpec Little Red Dot and Non─Little Red Dot Broad H{\ensuremath{\alpha}} Emitters in Light of Nondetection of NIRCam Photometric Variability and X-Ray}",
      journal = {\apj},
         year = 2025,
        month = dec,
       volume = {995},
       number = {1},
          eid = {24},
        pages = {24},
          doi = {10.3847/1538-4357/ae119e},
archivePrefix = {arXiv},
       eprint = {2407.04777},
 primaryClass = {astro-ph.GA},
       adsurl = {https://ui.adsabs.harvard.edu/abs/2025ApJ...995...24K}
}

@ARTICLE{Wang2024a,
       author = {{Wang}, Bingjie and {de Graaff}, Anna and {Davies}, Rebecca L. and {Greene}, Jenny E. and {Leja}, Joel and {Brammer}, Gabriel B. and {Goulding}, Andy D. and {Miller}, Tim B. and {Suess}, Katherine A. and {Weibel}, Andrea and {Williams}, Christina C. and {Bezanson}, Rachel and {Boogaard}, Leindert A. and {Cleri}, Nikko J. and {Hirschmann}, Michaela and {Katz}, Harley and {Labb{\'e}}, Ivo and {Maseda}, Michael V. and {Matthee}, Jorryt and {McConachie}, Ian and {Naidu}, Rohan P. and {Oesch}, Pascal A. and {Rix}, Hans-Walter and {Setton}, David J. and {Whitaker}, Katherine E.},
        title = "{RUBIES: JWST/NIRSpec Confirmation of an Infrared-luminous, Broad-line Little Red Dot with an Ionized Outflow}",
      journal = {\apj},
         year = 2025,
        month = may,
       volume = {984},
       number = {2},
          eid = {121},
        pages = {121},
          doi = {10.3847/1538-4357/adc1ca},
archivePrefix = {arXiv},
       eprint = {2403.02304},
 primaryClass = {astro-ph.GA},
       adsurl = {https://ui.adsabs.harvard.edu/abs/2025ApJ...984..121W}
}

@ARTICLE{Kokorev2024_break,
       author = {{Kokorev}, Vasily and {Chisholm}, John and {Endsley}, Ryan and {Finkelstein}, Steven L. and {Greene}, Jenny E. and {Akins}, Hollis B. and {Bromm}, Volker and {Casey}, Caitlin M. and {Fujimoto}, Seiji and {Labb{\'e}}, Ivo and {Larson}, Rebecca L.},
        title = "{Silencing the Giant: Evidence of Active Galactic Nucleus Feedback and Quenching in a Little Red Dot at z = 4.13}",
      journal = {\apj},
         year = 2024,
        month = nov,
       volume = {975},
       number = {2},
          eid = {178},
        pages = {178},
          doi = {10.3847/1538-4357/ad7d03},
archivePrefix = {arXiv},
       eprint = {2407.20320},
 primaryClass = {astro-ph.GA},
       adsurl = {https://ui.adsabs.harvard.edu/abs/2024ApJ...975..178K}
}

@ARTICLE{emcee,
       author = {{Foreman-Mackey}, Daniel and {Hogg}, David W. and {Lang}, Dustin and {Goodman}, Jonathan},
        title = "{emcee: The MCMC Hammer}",
      journal = {\pasp},
         year = 2013,
        month = mar,
       volume = {125},
       number = {925},
        pages = {306},
          doi = {10.1086/670067},
archivePrefix = {arXiv},
       eprint = {1202.3665},
 primaryClass = {astro-ph.IM},
       adsurl = {https://ui.adsabs.harvard.edu/abs/2013PASP..125..306F}
}

@ARTICLE{Zhang2024_variability,
       author = {{Zhang}, Zijian and {Jiang}, Linhua and {Liu}, Weiyang and {Ho}, Luis C.},
        title = "{Analysis of Multi-epoch JWST Images of {\ensuremath{\sim}}300 Little Red Dots: Tentative Detection of Variability in a Minority of Sources}",
      journal = {\apj},
         year = 2025,
        month = may,
       volume = {985},
       number = {1},
          eid = {119},
        pages = {119},
          doi = {10.3847/1538-4357/adcb3e},
archivePrefix = {arXiv},
       eprint = {2411.02729},
 primaryClass = {astro-ph.GA},
       adsurl = {https://ui.adsabs.harvard.edu/abs/2025ApJ...985..119Z}
}

@ARTICLE{Perger2024,
       author = {{Perger}, K. and {Fogasy}, J. and {Frey}, S. and {Gab{\'a}nyi}, K. {\'E}.},
        title = "{Deep silence: Radio properties of little red dots}",
      journal = {\aap},
         year = 2025,
        month = jan,
       volume = {693},
          eid = {L2},
        pages = {L2},
          doi = {10.1051/0004-6361/202452422},
archivePrefix = {arXiv},
       eprint = {2411.19518},
 primaryClass = {astro-ph.GA},
       adsurl = {https://ui.adsabs.harvard.edu/abs/2025A&A...693L...2P}
}

@ARTICLE{mazzolari2024,
       author = {{Mazzolari}, G. and {Gilli}, R. and {Maiolino}, R. and {Prandoni}, I. and {Delvecchio}, I. and {Norman}, C. and {Jim{\'e}nez-Andrade}, E.~F. and {Belladitta}, S. and {Vito}, F. and {Momjian}, E. and {Chiaberge}, M. and {Trefoloni}, B. and {Signorini}, M. and {Ji}, X. and {D'Amato}, Q. and {Risaliti}, G. and {Baldi}, R.~D. and {Fabian}, A. and {{\"U}bler}, H. and {D'Eugenio}, F. and {Scholtz}, J. and {Juod{\v{z}}balis}, I. and {Mignoli}, M. and {Brusa}, M. and {Murphy}, E.~J. and {Muxlow}, T.~W.~B.},
        title = "{The radio properties of the JWST-discovered AGN}",
      journal = {\aap},
         year = 2026,
        month = feb,
       volume = {706},
          eid = {A372},
        pages = {A372},
          doi = {10.1051/0004-6361/202453317},
archivePrefix = {arXiv},
       eprint = {2412.04224},
 primaryClass = {astro-ph.GA},
       adsurl = {https://ui.adsabs.harvard.edu/abs/2026A&A...706A.372M}
}

@ARTICLE{tee2024_variability,
       author = {{Tee}, Wei Leong and {Fan}, Xiaohui and {Wang}, Feige and {Yang}, Jinyi},
        title = "{Lack of Rest-frame Ultraviolet Variability in Little Red Dots Based on HST and JWST Observations}",
      journal = {\apjl},
         year = 2025,
        month = apr,
       volume = {983},
       number = {1},
          eid = {L26},
        pages = {L26},
          doi = {10.3847/2041-8213/adc5e3},
archivePrefix = {arXiv},
       eprint = {2412.05242},
 primaryClass = {astro-ph.GA},
       adsurl = {https://ui.adsabs.harvard.edu/abs/2025ApJ...983L..26T}
}

@ARTICLE{Tanaka2025_z10,
       author = {{Tanaka}, Takumi S. and {Akins}, Hollis B. and {Harikane}, Yuichi and {Silverman}, John D. and {Casey}, Caitlin M. and {Inayoshi}, Kohei and {Schindler}, Jan-Torge and {Shimasaku}, Kazuhiro and {Kocevski}, Dale D. and {Onoue}, Masafusa and {Faisst}, Andreas L. and {Robertson}, Brant E. and {Kokorev}, Vasily and {Shuntov}, Marko and {Koekemoer}, Anton M. and {Franco}, Maximilien and {Egami}, Eiichi and {Liu}, Daizhong and {Taylor}, Anthony J. and {Kartaltepe}, Jeyhan S. and {Bosman}, Sarah E.~I. and {Champagne}, Jaclyn B. and {Kakiichi}, Koki and {Harish}, Santosh and {Zhang}, Zijian and {Newman}, Sophie L. and {Kakkad}, Darshan and {Fei}, Qinyue and {Fujimoto}, Seiji and {Li}, Mingyu and {Finkelstein}, Steven L. and {Li}, Zi-Jian and {Lambrides}, Erini and {Sommovigo}, Laura and {Zavala}, Jorge A. and {Ito}, Kei and {Liu}, Zhaoxuan and {Treister}, Ezequiel and {Aravena}, Manuel and {Gozaliasl}, Ghassem and {Zhang}, Haowen and {Hatamnia}, Hossein and {Umeda}, Hiroya and {Inoue}, Akio K. and {Yang}, Jinyi and {Ando}, Makoto and {Arita}, Junya and {Ding}, Xuheng and {Matsui}, Suin and {Shibanuma}, Yuki and {Magdis}, Georgios and {Zhuang}, Mingyang and {Fan}, Xiaohui and {Li}, Zihao and {Liu}, Weizhe and {Lyu}, Jianwei and {Rhodes}, Jason and {Toft}, Sune and {Wang}, Feige and {Zou}, Siwei and {Arango-Toro}, Rafael C. and {Battisti}, Andrew J. and {Gillman}, Steven and {Khostovan}, Ali Ahmad and {Long}, Arianna S. and {Mobasher}, Bahram and {Sanders}, David B.},
        title = "{Discovery of a Little Red Dot Candidate at z {\ensuremath{\gtrsim}} 10 in COSMOS-web Based on MIRI-NIRCam Selection}",
      journal = {\apj},
         year = 2025,
        month = dec,
       volume = {995},
       number = {1},
          eid = {21},
        pages = {21},
          doi = {10.3847/1538-4357/ae145f},
archivePrefix = {arXiv},
       eprint = {2508.00057},
 primaryClass = {astro-ph.GA},
       adsurl = {https://ui.adsabs.harvard.edu/abs/2025ApJ...995...21T}
}

@ARTICLE{Rinaldi2024,
       author = {{Rinaldi}, P. and {Bonaventura}, N. and {Rieke}, G.~H. and {Alberts}, S. and {Caputi}, K.~I. and {Baker}, W.~M. and {Baum}, S. and {Bhatawdekar}, R. and {Bunker}, A.~J. and {Carniani}, S. and {Curtis-Lake}, E. and {D'Eugenio}, F. and {Egami}, E. and {Ji}, Z. and {Hainline}, K. and {Helton}, J.~M. and {Lin}, X. and {Lyu}, J. and {Johnson}, B.~D. and {Ma}, Z. and {Maiolino}, R. and {P{\'e}rez-Gonz{\'a}lez}, P.~G. and {Rieke}, M. and {Robertson}, B.~E. and {Shivaei}, I. and {Stone}, M. and {Sun}, Y. and {Tacchella}, S. and {{\"U}bler}, H. and {Williams}, C.~C. and {Willmer}, C.~N.~A. and {Willott}, C. and {Zhang}, J. and {Zhu}, Y.},
        title = "{Not Just a Dot: the complex UV morphology and underlying properties of Little Red Dots}",
      journal = {arXiv e-prints},
         year = 2024,
        month = nov,
          eid = {arXiv:2411.14383},
        pages = {arXiv:2411.14383},
          doi = {10.48550/arXiv.2411.14383},
archivePrefix = {arXiv},
       eprint = {2411.14383},
 primaryClass = {astro-ph.GA},
       adsurl = {https://ui.adsabs.harvard.edu/abs/2024arXiv241114383R}
}

@ARTICLE{Rusakov2025,
       author = {{Rusakov}, V. and {Watson}, D. and {Nikopoulos}, G.~P. and {Brammer}, G. and {Gottumukkala}, R. and {Harvey}, T. and {Heintz}, K.~E. and {Nielsen}, R.~D. and {Sim}, S.~A. and {Sneppen}, A. and {Vijayan}, A.~P. and {Adams}, N. and {Austin}, D. and {Conselice}, C.~J. and {Goolsby}, C.~M. and {Toft}, S.},
        title = "{JWST's little red dots: an emerging population of young, low-mass AGN cocooned in dense ionized gas}",
      journal = {arXiv e-prints},
         year = 2025,
        month = mar,
          eid = {arXiv:2503.16595},
        pages = {arXiv:2503.16595},
          doi = {10.48550/arXiv.2503.16595},
archivePrefix = {arXiv},
       eprint = {2503.16595},
 primaryClass = {astro-ph.GA},
       adsurl = {https://ui.adsabs.harvard.edu/abs/2025arXiv250316595R}
}

@ARTICLE{Juodzbalis2025,
       author = {{Juod{\v{z}}balis}, Ignas and {Maiolino}, Roberto and {Baker}, William M. and {Lake}, Emma Curtis and {Scholtz}, Jan and {D'Eugenio}, Francesco and {Trefoloni}, Bartolomeo and {Isobe}, Yuki and {Tacchella}, Sandro and {Bunker}, Andrew J. and {Carniani}, Stefano and {Charlot}, St{\'e}phane and {Jones}, Gareth C. and {Parlanti}, Eleonora and {Perna}, Michele and {Rinaldi}, Pierluigi and {Robertson}, Brant and {{\"U}bler}, Hannah and {Venturi}, Giacomo and {Willott}, Chris},
        title = "{JADES: comprehensive census of broad-line AGN from Reionization to Cosmic Noon revealed by JWST}",
      journal = {arXiv e-prints},
         year = 2025,
        month = apr,
          eid = {arXiv:2504.03551},
        pages = {arXiv:2504.03551},
          doi = {10.48550/arXiv.2504.03551},
archivePrefix = {arXiv},
       eprint = {2504.03551},
 primaryClass = {astro-ph.GA},
       adsurl = {https://ui.adsabs.harvard.edu/abs/2025arXiv250403551J}
}

@ARTICLE{Inayoshi2025,
       author = {{Inayoshi}, Kohei},
        title = "{Little Red Dots as the Very First Activity of Black Hole Growth}",
      journal = {\apjl},
         year = 2025,
        month = jul,
       volume = {988},
       number = {1},
          eid = {L22},
        pages = {L22},
          doi = {10.3847/2041-8213/adea66},
archivePrefix = {arXiv},
       eprint = {2503.05537},
 primaryClass = {astro-ph.GA},
       adsurl = {https://ui.adsabs.harvard.edu/abs/2025ApJ...988L..22I}
}

@ARTICLE{Hviding2026,
       author = {{Hviding}, Raphael E. and {de Graaff}, Anna and {Liu}, Hanpu and {Goulding}, Andy D. and {Ma}, Yilun and {Greene}, Jenny E. and {Boogaard}, Leindert A. and {Bunker}, Andrew J. and {Cleri}, Nikko J. and {Franx}, Marijn and {Hirschmann}, Michaela and {Leja}, Joel and {Matthee}, Jorryt and {Naidu}, Rohan P. and {Setton}, David J. and {{\"U}bler}, Hannah and {Venturi}, Giacomo and {Wang}, Bingjie},
        title = "{The X-Ray Dot: Exotic Dust or a Late-stage Little Red Dot?}",
      journal = {\apjl},
         year = 2026,
        month = mar,
       volume = {1000},
       number = {1},
          eid = {L18},
        pages = {L18},
          doi = {10.3847/2041-8213/ae4c88},
archivePrefix = {arXiv},
       eprint = {2601.09778},
 primaryClass = {astro-ph.GA},
       adsurl = {https://ui.adsabs.harvard.edu/abs/2026ApJ..1000L..18H}
}

@ARTICLE{Fu2026,
       author = {{Fu}, Shuqi and {Zhang}, Zijian and {Jiang}, Danyang and {Chen}, Jie and {Jiang}, Linhua and {Ho}, Luis C. and {Inayoshi}, Kohei and {Chen}, Kaiyuan and {Lyu}, Jianwei and {Sun}, Fengwu and {Wang}, Feige and {Yang}, Jinyi},
        title = "{The discovery of two little red dots in transition into quasars}",
      journal = {Nature Astronomy},
         year = 2026,
        month = sep,
       volume = {10},
       number = {9},
        pages = {1372-1386},
          doi = {10.1038/s41550-026-02885-8},
archivePrefix = {arXiv},
       eprint = {2512.02096},
 primaryClass = {astro-ph.GA},
       adsurl = {https://ui.adsabs.harvard.edu/abs/2026NatAs..10.1372F}
}

@ARTICLE{Rigby2023_JWST,
       author = {{Rigby}, Jane and {Perrin}, Marshall and {McElwain}, Michael and {Kimble}, Randy and {Friedman}, Scott and {Lallo}, Matt and {Doyon}, Ren{\'e} and {Feinberg}, Lee and {Ferruit}, Pierre and {Glasse}, Alistair and {Rieke}, Marcia and {Rieke}, George and {Wright}, Gillian and {Willott}, Chris and {Colon}, Knicole and {Milam}, Stefanie and {Neff}, Susan and {Stark}, Christopher and {Valenti}, Jeff and {Abell}, Jim and {Abney}, Faith and {Abul-Huda}, Yasin and {Acton}, D. Scott and {Adams}, Evan and {Adler}, David and {Aguilar}, Jonathan and {Ahmed}, Nasif and {Albert}, Lo{\"\i}c and {Alberts}, Stacey and {Aldridge}, David and {Allen}, Marsha and {Altenburg}, Martin and {{\'A}lvarez-M{\'a}rquez}, Javier and {Alves de Oliveira}, Catarina and {Andersen}, Greg and {Anderson}, Harry and {Anderson}, Sara and {Argyriou}, Ioannis and {Armstrong}, Amber and {Arribas}, Santiago and {Artigau}, Etienne and {Arvai}, Amanda and {Atkinson}, Charles and {Bacon}, Gregory and {Bair}, Thomas and {Banks}, Kimberly and {Barrientes}, Jaclyn and {Barringer}, Bruce and {Bartosik}, Peter and {Bast}, William and {Baudoz}, Pierre and {Beatty}, Thomas and {Bechtold}, Katie and {Beck}, Tracy and {Bergeron}, Eddie and {Bergkoetter}, Matthew and {Bhatawdekar}, Rachana and {Birkmann}, Stephan and {Blazek}, Ronald and {Blome}, Claire and {Boccaletti}, Anthony and {B{\"o}ker}, Torsten and {Boia}, John and {Bonaventura}, Nina and {Bond}, Nicholas and {Bosley}, Kari and {Boucarut}, Ray and {Bourque}, Matthew and {Bouwman}, Jeroen and {Bower}, Gary and {Bowers}, Charles and {Boyer}, Martha and {Bradley}, Larry and {Brady}, Greg and {Braun}, Hannah and {Breda}, David and {Bresnahan}, Pamela and {Bright}, Stacey and {Britt}, Christopher and {Bromenschenkel}, Asa and {Brooks}, Brian and {Brooks}, Keira and {Brown}, Bob and {Brown}, Matthew and {Brown}, Patricia and {Bunker}, Andy and {Burger}, Matthew and {Bushouse}, Howard and {Cale}, Steven and {Cameron}, Alex and {Cameron}, Peter and {Canipe}, Alicia and {Caplinger}, James and {Caputo}, Francis and {Cara}, Mihai and {Carey}, Larkin and {Carniani}, Stefano and {Carrasquilla}, Maria and {Carruthers}, Margaret and {Case}, Michael and {Catherine}, Riggs and {Chance}, Don and {Chapman}, George and {Charlot}, St{\'e}phane and {Charlow}, Brian and {Chayer}, Pierre and {Chen}, Bin and {Cherinka}, Brian and {Chichester}, Sarah and {Chilton}, Zack and {Chonis}, Taylor and {Clampin}, Mark and {Clark}, Charles and {Clark}, Kerry and {Coe}, Dan and {Coleman}, Benee and {Comber}, Brian and {Comeau}, Tom and {Connolly}, Dennis and {Cooper}, James and {Cooper}, Rachel and {Coppock}, Eric and {Correnti}, Matteo and {Cossou}, Christophe and {Coulais}, Alain and {Coyle}, Laura and {Cracraft}, Misty and {Curti}, Mirko and {Cuturic}, Steven and {Davis}, Katherine and {Davis}, Michael and {Dean}, Bruce and {DeLisa}, Amy and {deMeester}, Wim and {Dencheva}, Nadia and {Dencheva}, Nadezhda and {DePasquale}, Joseph and {Deschenes}, Jeremy and {Hunor Detre}, {\"O}rs and {Diaz}, Rosa and {Dicken}, Dan and {DiFelice}, Audrey and {Dillman}, Matthew and {Dixon}, William and {Doggett}, Jesse and {Donaldson}, Tom and {Douglas}, Rob and {DuPrie}, Kimberly and {Dupuis}, Jean and {Durning}, John and {Easmin}, Nilufar and {Eck}, Weston and {Edeani}, Chinwe and {Egami}, Eiichi and {Ehrenwinkler}, Ralf and {Eisenhamer}, Jonathan and {Eisenhower}, Michael and {Elie}, Michelle and {Elliott}, James and {Elliott}, Kyle and {Ellis}, Tracy and {Engesser}, Michael and {Espinoza}, Nestor and {Etienne}, Odessa and {Etxaluze}, Mireya and {Falini}, Patrick and {Feeney}, Matthew and {Ferry}, Malcolm and {Filippazzo}, Joseph and {Fincham}, Brian and {Fix}, Mees and {Flagey}, Nicolas and {Florian}, Michael and {Flynn}, Jim and {Fontanella}, Erin and {Ford}, Terrance and {Forshay}, Peter and {Fox}, Ori and {Franz}, David and {Fu}, Henry and {Fullerton}, Alexander and {Galkin}, Sergey and {Galyer}, Anthony and {Garc{\'\i}a Mar{\'\i}n}, Macarena and {Gardner}, Jonathan P. and {Gardner}, Lisa and {Garland}, Dennis and {Garrett}, Bruce and {Gasman}, Danny and {Gaspar}, Andras and {Gaudreau}, Daniel and {Gauthier}, Peter and {Geers}, Vincent and {Geithner}, Paul and {Gennaro}, Mario and {Giardino}, Giovanna and {Girard}, Julien and {Giuliano}, Mark and {Glassmire}, Kirk and {Glauser}, Adrian},
        title = "{The Science Performance of JWST as Characterized in Commissioning}",
      journal = {\pasp},
         year = 2023,
        month = apr,
       volume = {135},
       number = {1046},
          eid = {048001},
        pages = {048001},
          doi = {10.1088/1538-3873/acb293},
archivePrefix = {arXiv},
       eprint = {2207.05632},
 primaryClass = {astro-ph.IM},
       adsurl = {https://ui.adsabs.harvard.edu/abs/2023PASP..135d8001R}
}

@ARTICLE{Hviding2025,
       author = {{Hviding}, Raphael E. and {de Graaff}, Anna and {Miller}, Tim B. and {Setton}, David J. and {Greene}, Jenny E. and {Labb{\'e}}, Ivo and {Brammer}, Gabriel and {Bezanson}, Rachel and {Boogaard}, Leindert A. and {Cleri}, Nikko J. and {Leja}, Joel and {Maseda}, Michael V. and {McConachie}, Ian and {Matthee}, Jorryt and {Naidu}, Rohan P. and {Oesch}, Pascal A. and {Wang}, Bingjie and {Whitaker}, Katherine E. and {Williams}, Christina C.},
        title = "{RUBIES: A spectroscopic census of little red dots: All point sources with v-shaped continua have broad lines}",
      journal = {\aap},
         year = 2025,
        month = oct,
       volume = {702},
          eid = {A57},
        pages = {A57},
          doi = {10.1051/0004-6361/202555816},
archivePrefix = {arXiv},
       eprint = {2506.05459},
 primaryClass = {astro-ph.GA},
       adsurl = {https://ui.adsabs.harvard.edu/abs/2025A&A...702A..57H}
}

@ARTICLE{Lin2025_c3d,
       author = {{Lin}, Xiaojing and {Fan}, Xiaohui and {Wang}, Feige and {Sun}, Fengwu and {Champagne}, Jaclyn B. and {Egami}, Eiichi and {Kakiichi}, Koki and {Lyu}, Jianwei and {Tee}, Wei Leong and {Yang}, Jinyi and {Bian}, Fuyan and {Bosman}, Sarah E.~I. and {Cai}, Zheng and {Casey}, Caitlin M. and {Decarli}, Roberto and {Faisst}, Andreas L. and {Finkelstein}, Steven L. and {Fujimoto}, Seiji and {Harish}, Santosh and {Ilbert}, Olivier and {Inoue}, Akio K. and {Jin}, Xiangyu and {Kartaltepe}, Jeyhan S. and {Kocevski}, Dale D. and {Li}, Mingyu and {Liu}, Weizhe and {Liu}, Yichen and {Schindler}, Jan-Torge and {Shuntov}, Marko and {Tanaka}, Takumi S. and {Vestergaard}, Marianne and {Wu}, Yunjing and {Zhang}, Haowen and {Zhang}, Zijian},
        title = "{Bridging Quasars and Little Red Dots: Insights into Broad-line Active Galactic Nuclei at z = 5─8 from the First JWST COSMOS-3D Dataset}",
      journal = {\apj},
         year = 2026,
        month = jan,
       volume = {996},
       number = {1},
          eid = {93},
        pages = {93},
          doi = {10.3847/1538-4357/ae1b9b},
archivePrefix = {arXiv},
       eprint = {2504.08039},
 primaryClass = {astro-ph.GA},
       adsurl = {https://ui.adsabs.harvard.edu/abs/2026ApJ...996...93L}
}

@ARTICLE{Taylor2025,
       author = {{Taylor}, Anthony J. and {Kokorev}, Vasily and {Kocevski}, Dale D. and {Akins}, Hollis B. and {Cullen}, Fergus and {Dickinson}, Mark and {Finkelstein}, Steven L. and {Arrabal Haro}, Pablo and {Bromm}, Volker and {Giavalisco}, Mauro and {Inayoshi}, Kohei and {Juneau}, Stephanie and {Leung}, Gene C.~K. and {Perez-Gonzalez}, Pablo G. and {Somerville}, Rachel S. and {Trump}, Jonathan R. and {Amorin}, Ricardo O. and {Barro}, Guillermo and {Burgarella}, Denis and {Brooks}, Madisyn and {Carnall}, Adam and {Casey}, Caitlin M. and {Cheng}, Yingjie and {Chisholm}, John and {Chworowsky}, Katherine and {Davis}, Kelcey and {Donnan}, Callum T. and {Dunlop}, James S. and {Ellis}, Richard S. and {Fernandez}, Vital and {Fujimoto}, Seiji and {Grogin}, Norman A. and {Gupta}, Ansh R. and {Hathi}, Nimish P. and {Jung}, Intae and {Hirschmann}, Michaela and {Kartaltepe}, Jeyhan S. and {Koekemoer}, Anton M. and {Larson}, Rebecca L. and {Leung}, Ho-Hin and {Llerena}, Mario and {Lucas}, Ray A. and {McLeod}, Derek J. and {McLure}, Ross and {Napolitano}, Lorenzo and {Papovich}, Casey and {Stanton}, Thomas M. and {Tripodi}, Roberta and {Wang}, Xin and {Wilkins}, Stephen M. and {Yung}, L.~Y. Aaron and {Zavala}, Jorge A.},
        title = "{CAPERS-LRD-z9: A Gas Enshrouded Little Red Dot Hosting a Broad-line AGN at z=9.288}",
      journal = {arXiv e-prints},
         year = 2025,
        month = may,
          eid = {arXiv:2505.04609},
        pages = {arXiv:2505.04609},
          doi = {10.48550/arXiv.2505.04609},
archivePrefix = {arXiv},
       eprint = {2505.04609},
 primaryClass = {astro-ph.GA},
       adsurl = {https://ui.adsabs.harvard.edu/abs/2025arXiv250504609T}
}

@ARTICLE{Inoue2011,
       author = {{Inoue}, Akio K.},
        title = "{Rest-frame ultraviolet-to-optical spectral characteristics of extremely metal-poor and metal-free galaxies}",
      journal = {\mnras},
         year = 2011,
        month = aug,
       volume = {415},
       number = {3},
        pages = {2920-2931},
          doi = {10.1111/j.1365-2966.2011.18906.x},
archivePrefix = {arXiv},
       eprint = {1102.5150},
 primaryClass = {astro-ph.CO},
       adsurl = {https://ui.adsabs.harvard.edu/abs/2011MNRAS.415.2920I}
}

@ARTICLE{Withers2023,
       author = {{Withers}, Sunna and {Muzzin}, Adam and {Ravindranath}, Swara and {Sarrouh}, Ghassan T.~E. and {Abraham}, Roberto and {Asada}, Yoshihisa and {Brada{\v{c}}}, Maru{\v{s}}a and {Brammer}, Gabriel and {Desprez}, Guillaume and {Iyer}, Kartheik and {Martis}, Nicholas and {Mowla}, Lamiya and {Noirot}, Ga{\"e}l and {Sawicki}, Marcin and {Strait}, Victoria and {Willott}, Chris J.},
        title = "{Spectroscopy from Photometry: A Population of Extreme Emission Line Galaxies at 1.7 {\ensuremath{\lesssim}} z {\ensuremath{\lesssim}} 6.7 Selected with JWST Medium Band Filters}",
      journal = {\apjl},
         year = 2023,
        month = nov,
       volume = {958},
       number = {1},
          eid = {L14},
        pages = {L14},
          doi = {10.3847/2041-8213/ad01c0},
archivePrefix = {arXiv},
       eprint = {2304.11181},
 primaryClass = {astro-ph.GA},
       adsurl = {https://ui.adsabs.harvard.edu/abs/2023ApJ...958L..14W}
}

@ARTICLE{deGraaff2024_msa,
       author = {{de Graaff}, Anna and {Rix}, Hans-Walter and {Carniani}, Stefano and {Suess}, Katherine A. and {Charlot}, St{\'e}phane and {Curtis-Lake}, Emma and {Arribas}, Santiago and {Baker}, William M. and {Boyett}, Kristan and {Bunker}, Andrew J. and {Cameron}, Alex J. and {Chevallard}, Jacopo and {Curti}, Mirko and {Eisenstein}, Daniel J. and {Franx}, Marijn and {Hainline}, Kevin and {Hausen}, Ryan and {Ji}, Zhiyuan and {Johnson}, Benjamin D. and {Jones}, Gareth C. and {Maiolino}, Roberto and {Maseda}, Michael V. and {Nelson}, Erica and {Parlanti}, Eleonora and {Rawle}, Tim and {Robertson}, Brant and {Tacchella}, Sandro and {{\"U}bler}, Hannah and {Williams}, Christina C. and {Willmer}, Christopher N.~A. and {Willott}, Chris},
        title = "{Ionised gas kinematics and dynamical masses of z {\ensuremath{\gtrsim}} 6 galaxies from JADES/NIRSpec high-resolution spectroscopy}",
      journal = {\aap},
         year = 2024,
        month = apr,
       volume = {684},
          eid = {A87},
        pages = {A87},
          doi = {10.1051/0004-6361/202347755},
archivePrefix = {arXiv},
       eprint = {2308.09742},
 primaryClass = {astro-ph.GA},
       adsurl = {https://ui.adsabs.harvard.edu/abs/2024A&A...684A..87D}
}

@ARTICLE{deGraaff2025,
       author = {{de Graaff}, Anna and {Rix}, Hans-Walter and {Naidu}, Rohan P. and {Labb{\'e}}, Ivo and {Wang}, Bingjie and {Leja}, Joel and {Matthee}, Jorryt and {Katz}, Harley and {Greene}, Jenny E. and {Hviding}, Raphael E. and {Baggen}, Josephine and {Bezanson}, Rachel and {Boogaard}, Leindert A. and {Brammer}, Gabriel and {Dayal}, Pratika and {van Dokkum}, Pieter and {Goulding}, Andy D. and {Hirschmann}, Michaela and {Maseda}, Michael V. and {McConachie}, Ian and {Miller}, Tim B. and {Nelson}, Erica and {Oesch}, Pascal A. and {Setton}, David J. and {Shivaei}, Irene and {Weibel}, Andrea and {Whitaker}, Katherine E. and {Williams}, Christina C.},
        title = "{A remarkable ruby: Absorption in dense gas, rather than evolved stars, drives the extreme Balmer break of a little red dot at z = 3.5}",
      journal = {\aap},
         year = 2025,
        month = sep,
       volume = {701},
          eid = {A168},
        pages = {A168},
          doi = {10.1051/0004-6361/202554681},
archivePrefix = {arXiv},
       eprint = {2503.16600},
 primaryClass = {astro-ph.GA},
       adsurl = {https://ui.adsabs.harvard.edu/abs/2025A&A...701A.168D}
}

@ARTICLE{Naidu2025,
       author = {{Naidu}, Rohan P. and {Matthee}, Jorryt and {Katz}, Harley and {de Graaff}, Anna and {Oesch}, Pascal A. and {Smith}, Aaron and {Greene}, Jenny E. and {Brammer}, Gabriel and {Weibel}, Andrea and {Hviding}, Raphael and {Chisholm}, John and {Labb{\'e}}, Ivo and {Simcoe}, Robert A. and {Witten}, Callum and {Sun}, Wendy Q. and {Atek}, Hakim and {Baggen}, Josephine F.~W. and {Belli}, Sirio and {Bezanson}, Rachel and {Boogaard}, Leindert A. and {Bose}, Sownak and {Bouwens}, Rychard J. and {Covelo-Paz}, Alba and {Dayal}, Pratika and {Fudamoto}, Yoshinobu and {Furtak}, Lukas J. and {Giovinazzo}, Emma and {Goulding}, Andy and {Gronke}, Max and {Heintz}, Kasper E. and {Hirschmann}, Michaela and {Illingworth}, Garth and {Inoue}, Akio K. and {Johnson}, Benjamin D. and {Leja}, Joel and {Leonova}, Ecaterina and {McConachie}, Ian and {Maseda}, Michael V. and {Natarajan}, Priyamvada and {Nelson}, Erica and {Setton}, David J. and {Shivaei}, Irene and {Sobral}, David and {Stefanon}, Mauro and {Tacchella}, Sandro and {Toft}, Sune and {Torralba}, Alberto and {van Dokkum}, Pieter and {van der Wel}, Arjen and {Volonteri}, Marta and {Walter}, Fabian and {Wang}, Bingjie and {Watson}, Darach and {Whitaker}, Katherine},
        title = "{A gas-enshrouded and gas-reddened black hole at cosmic dawn}",
      journal = {\nat},
         year = 2026,
        month = aug,
       volume = {656},
       number = {8127},
        pages = {329-333},
          doi = {10.1038/s41586-026-10846-4},
archivePrefix = {arXiv},
       eprint = {2503.16596},
 primaryClass = {astro-ph.GA},
       adsurl = {https://ui.adsabs.harvard.edu/abs/2026Natur.656..329N}
}

@ARTICLE{Casey2025_dust,
       author = {{Casey}, Caitlin M. and {Akins}, Hollis B. and {Finkelstein}, Steven L. and {Franco}, Maximilien and {Fujimoto}, Seiji and {Liu}, Daizhong and {Long}, Arianna S. and {Magdis}, Georgios and {Manning}, Sinclaire M. and {McKinney}, Jed and {Shuntov}, Marko and {Tanaka}, Takumi S.},
        title = "{An upper limit of 10$^6$ M$_\odot$ in dust from ALMA observations in 60 Little Red Dots}",
      journal = {arXiv e-prints},
         year = 2025,
        month = may,
          eid = {arXiv:2505.18873},
        pages = {arXiv:2505.18873},
          doi = {10.48550/arXiv.2505.18873},
archivePrefix = {arXiv},
       eprint = {2505.18873},
 primaryClass = {astro-ph.GA},
       adsurl = {https://ui.adsabs.harvard.edu/abs/2025arXiv250518873C}
}

@ARTICLE{Setton2025,
       author = {{Setton}, David J. and {Greene}, Jenny E. and {Spilker}, Justin S. and {Williams}, Christina C. and {Labb{\'e}}, Ivo and {Ma}, Yilun and {Wang}, Bingjie and {Whitaker}, Katherine E. and {Leja}, Joel and {de Graaff}, Anna and {Alberts}, Stacey and {Bezanson}, Rachel and {Boogaard}, Leindert A. and {Brammer}, Gabriel and {Cutler}, Sam E. and {Cleri}, Nikko J. and {Cooper}, Olivia R. and {Dayal}, Pratika and {Fujimoto}, Seiji and {Furtak}, Lukas J. and {Goulding}, Andy D. and {Hirschmann}, Michaela and {Kokorev}, Vasily and {Maseda}, Michael V. and {McConachie}, Ian and {Matthee}, Jorryt and {Miller}, Tim B. and {Naidu}, Rohan P. and {Oesch}, Pascal A. and {Pan}, Richard and {Price}, Sedona H. and {Suess}, Katherine A. and {Weaver}, John R. and {Xiao}, Mengyuan and {Zhang}, Yunchong and {Zitrin}, Adi},
        title = "{A Confirmed Deficit of Hot and Cold Dust Emission in the Most Luminous Little Red Dots}",
      journal = {\apjl},
         year = 2025,
        month = sep,
       volume = {991},
       number = {1},
          eid = {L10},
        pages = {L10},
          doi = {10.3847/2041-8213/ade78b},
archivePrefix = {arXiv},
       eprint = {2503.02059},
 primaryClass = {astro-ph.GA},
       adsurl = {https://ui.adsabs.harvard.edu/abs/2025ApJ...991L..10S}
}

@ARTICLE{Volonteri2010,
       author = {{Volonteri}, Marta},
        title = "{Formation of supermassive black holes}",
      journal = {\aapr},
         year = 2010,
        month = jul,
       volume = {18},
       number = {3},
        pages = {279-315},
          doi = {10.1007/s00159-010-0029-x},
archivePrefix = {arXiv},
       eprint = {1003.4404},
 primaryClass = {astro-ph.CO},
       adsurl = {https://ui.adsabs.harvard.edu/abs/2010A&ARv..18..279V}
}

@ARTICLE{Setton2024,
       author = {{Setton}, David J. and {Greene}, Jenny E. and {de Graaff}, Anna and {Ma}, Yilun and {Leja}, Joel and {Matthee}, Jorryt and {Bezanson}, Rachel and {Boogaard}, Leindert A. and {Cleri}, Nikko J. and {Katz}, Harley and {Labbe}, Ivo and {Maseda}, Michael V. and {McConachie}, Ian and {Miller}, Tim B. and {Price}, Sedona H. and {Suess}, Katherine A. and {van Dokkum}, Pieter and {Wang}, Bingjie and {Weibel}, Andrea and {Whitaker}, Katherine E. and {Williams}, Christina C.},
        title = "{Little Red Dots at an Inflection Point: Ubiquitous V-shaped Turnover Consistently Occurs at the Balmer Limit}",
      journal = {\apj},
         year = 2025,
        month = dec,
       volume = {995},
       number = {1},
          eid = {118},
        pages = {118},
          doi = {10.3847/1538-4357/ae1500},
archivePrefix = {arXiv},
       eprint = {2411.03424},
 primaryClass = {astro-ph.GA},
       adsurl = {https://ui.adsabs.harvard.edu/abs/2025ApJ...995..118S}
}

@ARTICLE{Ji2025,
       author = {{Ji}, Xihan and {Maiolino}, Roberto and {{\"U}bler}, Hannah and {Scholtz}, Jan and {D'Eugenio}, Francesco and {Sun}, Fengwu and {Perna}, Michele and {Turner}, Hannah and {Carniani}, Stefano and {Arribas}, Santiago and {Bennett}, Jake S. and {Bunker}, Andrew and {Charlot}, St{\'e}phane and {Cresci}, Giovanni and {Curti}, Mirko and {Egami}, Eiichi and {Fabian}, Andy and {Inayoshi}, Kohei and {Isobe}, Yuki and {Jones}, Gareth and {Juod{\v{z}}balis}, Ignas and {Kumari}, Nimisha and {Lyu}, Jianwei and {Mazzolari}, Giovanni and {Parlanti}, Eleonora and {Robertson}, Brant and {Rodr{\'\i}guez Del Pino}, Bruno and {Schneider}, Raffaella and {Sijacki}, Debora and {Tacchella}, Sandro and {Trinca}, Alessandro and {Valiante}, Rosa and {Venturi}, Giacomo and {Volonteri}, Marta and {Willott}, Chris and {Witten}, Callum and {Witstok}, Joris},
        title = "{BlackTHUNDER ─ A non-stellar Balmer break in a black hole-dominated little red dot at z = 7.04}",
      journal = {\mnras},
         year = 2025,
        month = dec,
       volume = {544},
       number = {4},
        pages = {3900-3935},
          doi = {10.1093/mnras/staf1867},
archivePrefix = {arXiv},
       eprint = {2501.13082},
 primaryClass = {astro-ph.GA},
       adsurl = {https://ui.adsabs.harvard.edu/abs/2025MNRAS.544.3900J}
}

@ARTICLE{Zhou2025_variability,
       author = {{Zhou}, Shuying and {Sun}, Mouyuan and {Zhang}, Zijian and {Chen}, Jie and {Ho}, Luis C.},
        title = "{On the Variability Features of Active Galactic Nuclei in Little Red Dots}",
      journal = {\apj},
         year = 2025,
        month = oct,
       volume = {991},
       number = {2},
          eid = {137},
        pages = {137},
          doi = {10.3847/1538-4357/adfd5f},
archivePrefix = {arXiv},
       eprint = {2508.16795},
 primaryClass = {astro-ph.GA},
       adsurl = {https://ui.adsabs.harvard.edu/abs/2025ApJ...991..137Z}
}

@ARTICLE{Labbe2024,
       author = {{Labbe}, Ivo and {Greene}, Jenny E. and {Matthee}, Jorryt and {Treiber}, Helena and {Kokorev}, Vasily and {Miller}, Tim B. and {Kramarenko}, Ivan and {Setton}, David J. and {Ma}, Yilun and {Goulding}, Andy D. and {Bezanson}, Rachel and {Naidu}, Rohan P. and {Williams}, Christina C. and {Atek}, Hakim and {Brammer}, Gabriel and {Cutler}, Sam E. and {Chemerynska}, Iryna and {Cloonan}, Aidan P. and {Dayal}, Pratika and {de Graaff}, Anna and {Fudamoto}, Yoshinobu and {Fujimoto}, Seiji and {Furtak}, Lukas J. and {Glazebrook}, Karl and {Heintz}, Kasper E. and {Leja}, Joel and {Marchesini}, Danilo and {Nanayakkara}, Themiya and {Nelson}, Erica J. and {Oesch}, Pascal A. and {Pan}, Richard and {Price}, Sedona H. and {Shivaei}, Irene and {Sobral}, David and {Suess}, Katherine A. and {van Dokkum}, Pieter and {Wang}, Bingjie and {Weaver}, John R. and {Whitaker}, Katherine E. and {Zitrin}, Adi},
        title = "{An unambiguous AGN and a Balmer break in an Ultraluminous Little Red Dot at z=4.47 from Ultradeep UNCOVER and All the Little Things Spectroscopy}",
      journal = {arXiv e-prints},
         year = 2024,
        month = dec,
          eid = {arXiv:2412.04557},
        pages = {arXiv:2412.04557},
          doi = {10.48550/arXiv.2412.04557},
archivePrefix = {arXiv},
       eprint = {2412.04557},
 primaryClass = {astro-ph.GA},
       adsurl = {https://ui.adsabs.harvard.edu/abs/2024arXiv241204557L}
}

@ARTICLE{Kido2025,
       author = {{Kido}, Daisaburo and {Ioka}, Kunihito and {Hotokezaka}, Kenta and {Inayoshi}, Kohei and {Irwin}, Christopher M.},
        title = "{Black hole envelopes in Little Red Dots}",
      journal = {\mnras},
         year = 2025,
        month = dec,
       volume = {544},
       number = {4},
        pages = {3407-3416},
          doi = {10.1093/mnras/staf1898},
archivePrefix = {arXiv},
       eprint = {2505.06965},
 primaryClass = {astro-ph.HE},
       adsurl = {https://ui.adsabs.harvard.edu/abs/2025MNRAS.544.3407K}
}

@ARTICLE{Hainline2025,
       author = {{Hainline}, Kevin N. and {Maiolino}, Roberto and {Juod{\v{z}}balis}, Ignas and {Scholtz}, Jan and {{\"U}bler}, Hannah and {D'Eugenio}, Francesco and {Helton}, Jakob M. and {Sun}, Yang and {Sun}, Fengwu and {Robertson}, Brant and {Tacchella}, Sandro and {Bunker}, Andrew J. and {Carniani}, Stefano and {Charlot}, Stephane and {Curtis-Lake}, Emma and {Egami}, Eiichi and {Johnson}, Benjamin D. and {Lin}, Xiaojing and {Lyu}, Jianwei and {P{\'e}rez-Gonz{\'a}lez}, Pablo G. and {Rinaldi}, Pierluigi and {Silcock}, Maddie S. and {Venturi}, Giacomo and {Williams}, Christina C. and {Willmer}, Christopher N.~A. and {Willott}, Chris and {Zhang}, Junyu and {Zhu}, Yongda},
        title = "{An Investigation into the Selection and Colors of Little Red Dots and Active Galactic Nuclei}",
      journal = {\apj},
         year = 2025,
        month = feb,
       volume = {979},
       number = {2},
          eid = {138},
        pages = {138},
          doi = {10.3847/1538-4357/ad9920},
archivePrefix = {arXiv},
       eprint = {2410.00100},
 primaryClass = {astro-ph.GA},
       adsurl = {https://ui.adsabs.harvard.edu/abs/2025ApJ...979..138H}
}

@ARTICLE{GREXPLUS_SB,
       author = {{\!GREX-PLUS Science Team} and {Inoue}, Akio K. and {Harikane}, Yuichi and {Moriya}, Takashi and {Nomura}, Hideko and {Baba}, Shunsuke and {Fujii}, Yuka and {Gouda}, Naoteru and {Hirahara}, Yasuhiro and {Kawashima}, Yui and {Kodama}, Tadayuki and {Koyama}, Yusei and {Kurokawa}, Hiroyuki and {Matsuo}, Taro and {Matsuoka}, Yoshiki and {Matsuura}, Shuji and {Mawatari}, Ken and {Misawa}, Toru and {Nagamine}, Kentaro and {Nakajima}, Kimihiko and {Notsu}, Shota and {Ootsubo}, Takafumi and {Ohno}, Kazumasa and {Sagawa}, Hideo and {Shimonishi}, Takashi and {Tadaki}, Ken-ichi and {Takami}, Michihiro and {Terai}, Tsuyoshi and {Toba}, Yoshiki and {Yamashita}, Takuji and {Yasui}, Chikako},
        title = "{GREX-PLUS Science Book}",
      journal = {arXiv e-prints},
         year = 2023,
        month = apr,
          eid = {arXiv:2304.08104},
        pages = {arXiv:2304.08104},
          doi = {10.48550/arXiv.2304.08104},
archivePrefix = {arXiv},
       eprint = {2304.08104},
 primaryClass = {astro-ph.CO},
       adsurl = {https://ui.adsabs.harvard.edu/abs/2023arXiv230408104G}
}

@ARTICLE{Rodriguez2026,
       author = {{Rodriguez}, L.~F. and {Mirabel}, I.~F.},
        title = "{Radio detection of a local little red dot}",
      journal = {\aap},
         year = 2026,
        month = mar,
       volume = {707},
          eid = {L17},
        pages = {L17},
          doi = {10.1051/0004-6361/202558326},
archivePrefix = {arXiv},
       eprint = {2512.03331},
 primaryClass = {astro-ph.GA},
       adsurl = {https://ui.adsabs.harvard.edu/abs/2026A&A...707L..17R}
}

@ARTICLE{Rodriguez2026_2,
       author = {{Rodr{\'\i}guez}, Luis F. and {Mirabel}, I. F{\'e}lix and {Gonz{\'a}lez-L{\'o}pezlira}, Rosa A. and {Loinard}, Laurent},
        title = "{A Radio-bright Local Little Red Dot Analog}",
      journal = {\apj},
         year = 2026,
        month = oct,
       volume = {1009},
       number = {2},
          eid = {217},
        pages = {217},
          doi = {10.3847/1538-4357/ae9b01},
       adsurl = {https://ui.adsabs.harvard.edu/abs/2026ApJ..1009..217R}
}

@ARTICLE{Planck2018,
       author = {{\hspace{-3pt}Planck Collaboration} and {Aghanim}, N. and {Akrami}, Y. and {Ashdown}, M. and {Aumont}, J. and {Baccigalupi}, C. and {Ballardini}, M. and {Banday}, A.~J. and {Barreiro}, R.~B. and {Bartolo}, N. and {Basak}, S. and {Battye}, R. and {Benabed}, K. and {Bernard}, J.-P. and {Bersanelli}, M. and {Bielewicz}, P. and {Bock}, J.~J. and {Bond}, J.~R. and {Borrill}, J. and {Bouchet}, F.~R. and {Boulanger}, F. and {Bucher}, M. and {Burigana}, C. and {Butler}, R.~C. and {Calabrese}, E. and {Cardoso}, J.-F. and {Carron}, J. and {Challinor}, A. and {Chiang}, H.~C. and {Chluba}, J. and {Colombo}, L.~P.~L. and {Combet}, C. and {Contreras}, D. and {Crill}, B.~P. and {Cuttaia}, F. and {de Bernardis}, P. and {de Zotti}, G. and {Delabrouille}, J. and {Delouis}, J.-M. and {Di Valentino}, E. and {Diego}, J.~M. and {Dor{\'e}}, O. and {Douspis}, M. and {Ducout}, A. and {Dupac}, X. and {Dusini}, S. and {Efstathiou}, G. and {Elsner}, F. and {En{\ss}lin}, T.~A. and {Eriksen}, H.~K. and {Fantaye}, Y. and {Farhang}, M. and {Fergusson}, J. and {Fernandez-Cobos}, R. and {Finelli}, F. and {Forastieri}, F. and {Frailis}, M. and {Fraisse}, A.~A. and {Franceschi}, E. and {Frolov}, A. and {Galeotta}, S. and {Galli}, S. and {Ganga}, K. and {G{\'e}nova-Santos}, R.~T. and {Gerbino}, M. and {Ghosh}, T. and {Gonz{\'a}lez-Nuevo}, J. and {G{\'o}rski}, K.~M. and {Gratton}, S. and {Gruppuso}, A. and {Gudmundsson}, J.~E. and {Hamann}, J. and {Handley}, W. and {Hansen}, F.~K. and {Herranz}, D. and {Hildebrandt}, S.~R. and {Hivon}, E. and {Huang}, Z. and {Jaffe}, A.~H. and {Jones}, W.~C. and {Karakci}, A. and {Keih{\"a}nen}, E. and {Keskitalo}, R. and {Kiiveri}, K. and {Kim}, J. and {Kisner}, T.~S. and {Knox}, L. and {Krachmalnicoff}, N. and {Kunz}, M. and {Kurki-Suonio}, H. and {Lagache}, G. and {Lamarre}, J.-M. and {Lasenby}, A. and {Lattanzi}, M. and {Lawrence}, C.~R. and {Le Jeune}, M. and {Lemos}, P. and {Lesgourgues}, J. and {Levrier}, F. and {Lewis}, A. and {Liguori}, M. and {Lilje}, P.~B. and {Lilley}, M. and {Lindholm}, V. and {L{\'o}pez-Caniego}, M. and {Lubin}, P.~M. and {Ma}, Y.-Z. and {Mac{\'\i}as-P{\'e}rez}, J.~F. and {Maggio}, G. and {Maino}, D. and {Mandolesi}, N. and {Mangilli}, A. and {Marcos-Caballero}, A. and {Maris}, M. and {Martin}, P.~G. and {Martinelli}, M. and {Mart{\'\i}nez-Gonz{\'a}lez}, E. and {Matarrese}, S. and {Mauri}, N. and {McEwen}, J.~D. and {Meinhold}, P.~R. and {Melchiorri}, A. and {Mennella}, A. and {Migliaccio}, M. and {Millea}, M. and {Mitra}, S. and {Miville-Desch{\^e}nes}, M.-A. and {Molinari}, D. and {Montier}, L. and {Morgante}, G. and {Moss}, A. and {Natoli}, P. and {N{\o}rgaard-Nielsen}, H.~U. and {Pagano}, L. and {Paoletti}, D. and {Partridge}, B. and {Patanchon}, G. and {Peiris}, H.~V. and {Perrotta}, F. and {Pettorino}, V. and {Piacentini}, F. and {Polastri}, L. and {Polenta}, G. and {Puget}, J.-L. and {Rachen}, J.~P. and {Reinecke}, M. and {Remazeilles}, M. and {Renzi}, A. and {Rocha}, G. and {Rosset}, C. and {Roudier}, G. and {Rubi{\~n}o-Mart{\'\i}n}, J.~A. and {Ruiz-Granados}, B. and {Salvati}, L. and {Sandri}, M. and {Savelainen}, M. and {Scott}, D. and {Shellard}, E.~P.~S. and {Sirignano}, C. and {Sirri}, G. and {Spencer}, L.~D. and {Sunyaev}, R. and {Suur-Uski}, A.-S. and {Tauber}, J.~A. and {Tavagnacco}, D. and {Tenti}, M. and {Toffolatti}, L. and {Tomasi}, M. and {Trombetti}, T. and {Valenziano}, L. and {Valiviita}, J. and {Van Tent}, B. and {Vibert}, L. and {Vielva}, P. and {Villa}, F. and {Vittorio}, N. and {Wandelt}, B.~D. and {Wehus}, I.~K. and {White}, M. and {White}, S.~D.~M. and {Zacchei}, A. and {Zonca}, A.},
        title = "{Planck 2018 results. VI. Cosmological parameters}",
      journal = {\aap},
         year = 2020,
        month = sep,
       volume = {641},
          eid = {A6},
        pages = {A6},
          doi = {10.1051/0004-6361/201833910},
archivePrefix = {arXiv},
       eprint = {1807.06209},
 primaryClass = {astro-ph.CO},
       adsurl = {https://ui.adsabs.harvard.edu/abs/2020A&A...641A...6P}
}

@ARTICLE{Sun2026,
       author = {{Sun}, Wendy Q. and {Naidu}, Rohan P. and {Matthee}, Jorryt and {de Graaff}, Anna and {Chisholm}, John and {Greene}, Jenny E. and {Oesch}, Pascal A. and {Torralba}, Alberto and {Hviding}, Raphael E. and {Brammer}, Gabriel and {Simcoe}, Robert A. and {Bose}, Sownak and {Bouwens}, Rychard and {Dayal}, Pratika and {Eilers}, Anna-Christina and {Fei}, Qinyue and {Furtak}, Lukas J. and {Gottumukkala}, Rashmi and {Goulding}, Andy and {Heintz}, Kasper E. and {Hirschmann}, Michaela and {Kokorev}, Vasily and {Leja}, Joel and {Liu}, Zhaoran and {Natarajan}, Priyamvada and {Santarelli}, Andrew D. and {Setton}, David J. and {Smith}, Aaron and {Tacchella}, Sandro and {Volonteri}, Marta and {Walter}, Fabian and {Weibel}, Andrea and {Williams}, Christina C.},
        title = "{Little Red Dot {\ensuremath{-}} Host Galaxy = Black Hole Star: A Gas-Enshrouded Heart at the Center of Every Little Red Dot}",
      journal = {The Open Journal of Astrophysics},
         year = 2026,
        month = may,
       volume = {9},
        pages = {62505},
          doi = {10.33232/001c.162505},
archivePrefix = {arXiv},
       eprint = {2601.20929},
 primaryClass = {astro-ph.GA},
       adsurl = {https://ui.adsabs.harvard.edu/abs/2026OJAp....962505S}
}

@INPROCEEDINGS{GREXPLUS_SPIE2024,
       author = {{Inoue}, Akio K. and {Yamamura}, Issei and {Suzuki}, Toyoaki and {Nakagawa}, Takao and {Kaneda}, Hidehiro and {Nomura}, Hideko and {Kodama}, Tadayuki and {Wada}, Takehiko and {Iwamuro}, Fumihide and {Motohara}, Kentaro and {Komiyama}, Yutaka and {Oyabu}, Shinki and {Harikane}, Yuichi and {Moriya}, Takashi and {Ouchi}, Masami and {Yamada}, Toru and {Notsu}, Shota},
        title = "{Concept study of GREX-PLUS: galaxy reionization explorer and planetary universe spectrometer}",
    booktitle = {Space Telescopes and Instrumentation 2024: Optical, Infrared, and Millimeter Wave},
         year = 2024,
       editor = {{Coyle}, Laura E. and {Matsuura}, Shuji and {Perrin}, Marshall D.},
       series = {Society of Photo-Optical Instrumentation Engineers (SPIE) Conference Series},
       volume = {13092},
        month = aug,
          eid = {130920Y},
        pages = {130920Y},
          doi = {10.1117/12.3017780},
       adsurl = {https://ui.adsabs.harvard.edu/abs/2024SPIE13092E..0YI}
}

@ARTICLE{Furtak2025_variability,
       author = {{Furtak}, Lukas J. and {Secunda}, Amy R. and {Greene}, Jenny E. and {Zitrin}, Adi and {Labb{\'e}}, Ivo and {Golubchik}, Miriam and {Bezanson}, Rachel and {Kokorev}, Vasily and {Atek}, Hakim and {Brammer}, Gabriel B. and {Chemerynska}, Iryna and {Cutler}, Sam E. and {Dayal}, Pratika and {Feldmann}, Robert and {Fujimoto}, Seiji and {Glazebrook}, Karl and {Leja}, Joel and {Ma}, Yilun and {Matthee}, Jorryt and {Naidu}, Rohan P. and {Nelson}, Erica J. and {Oesch}, Pascal A. and {Pan}, Richard and {Price}, Sedona H. and {Suess}, Katherine A. and {Wang}, Bingjie and {Weaver}, John R. and {Whitaker}, Katherine E.},
        title = "{Investigating photometric and spectroscopic variability in the multiply imaged little red dot A2744-QSO1}",
      journal = {\aap},
         year = 2025,
        month = jun,
       volume = {698},
          eid = {A227},
        pages = {A227},
          doi = {10.1051/0004-6361/202554110},
archivePrefix = {arXiv},
       eprint = {2502.07875},
 primaryClass = {astro-ph.GA},
       adsurl = {https://ui.adsabs.harvard.edu/abs/2025A&A...698A.227F}
}

@ARTICLE{Zhang2025_EinsteinCross,
       author = {{Zhang}, Zijian and {Li}, Mingyu and {Oguri}, Masamune and {Lin}, Xiaojing and {Inayoshi}, Kohei and {Cerny}, Catherine and {Coe}, Dan and {Diego}, Jose M. and {Fujimoto}, Seiji and {Jiang}, Linhua and {Mahler}, Guillaume and {Matthee}, Jorryt and {Naidu}, Rohan P. and {Sharon}, Keren and {Shen}, Yue and {Zitrin}, Adi and {Abdurro'uf} and {Akins}, Hollis and {Allingham}, Joseph F.~V. and {Amor{\'\i}n}, Ricardo and {Asada}, Yoshihisa and {Atek}, Hakim and {Bauer}, Franz E. and {Brada{\v{c}}}, Maru{\v{s}}a and {Bradley}, Larry D. and {Cai}, Zheng and {Cantalupo}, Sebastiano and {Conselice}, Christopher and {Dai}, Liang and {Dayal}, Pratika and {Egami}, Eiichi and {Eisenstein}, Daniel J. and {Faisst}, Andreas L. and {Fan}, Xiaohui and {Fei}, Qinyue and {Frye}, Brenda L. and {Fudamoto}, Yoshinobu and {Furtak}, Lukas J. and {Golubchik}, Miriam and {Gonz{\'a}lez-Otero}, Mauro and {Harikane}, Yuichi and {Hsiao}, Tiger Yu-Yang and {Jim{\'e}nez-Teja}, Yolanda and {Kartaltepe}, Jeyhan S. and {Kiyota}, Tomokazu and {Koekemoer}, Anton M. and {Kohno}, Kotaro and {Kokorev}, Vasily and {Kumari}, Nimisha and {Labbe}, Ivo and {Lagos}, Claudia D.~P. and {Larison}, Conor and {Liang}, Yongming and {Lucas}, Ray A. and {Lyu}, Jianwei and {Martis}, Nicholas S. and {Magdis}, Georgios E. and {Messa}, Matteo and {Nakane}, Minami and {Noirot}, Ga{\"e}l and {Ortiz}, III, Rafael and {Ouchi}, Masami and {Pierel}, Justin D.~R. and {Postman}, Marc and {Reddy}, Naveen and {Ricotti}, Massimo and {Schaerer}, Daniel and {Schneider}, Raffaella and {Steidel}, Charles C. and {Tee}, Wei Leong and {Tripodi}, Roberta and {Trussler}, James A.~A. and {Umeda}, Hiroya and {Valentino}, Francesco and {Vanzella}, Eros and {Wang}, Feige and {Windhorst}, Rogier and {Wu}, Yunjing and {Wu}, Zihao and {Yanagisawa}, Hiroto and {Yang}, Jinyi and {Sun}, Fengwu},
        title = "{Little red dot variability over a century reveals black hole envelope via a giant Einstein cross}",
      journal = {arXiv e-prints},
         year = 2025,
        month = dec,
          eid = {arXiv:2512.05180},
        pages = {arXiv:2512.05180},
          doi = {10.48550/arXiv.2512.05180},
archivePrefix = {arXiv},
       eprint = {2512.05180},
 primaryClass = {astro-ph.GA},
       adsurl = {https://ui.adsabs.harvard.edu/abs/2025arXiv251205180Z}
}

@ARTICLE{Inayoshi2025_seduni,
       author = {{Inayoshi}, Kohei and {Murase}, Kohta and {Kashiyama}, Kazumi},
        title = "{Spectral Uniformity of Little Red Dots: A Natural Outcome of Coevolving Seed Black Holes and Nascent Starbursts}",
      journal = {\apj},
         year = 2026,
        month = mar,
       volume = {1000},
       number = {1},
          eid = {90},
        pages = {90},
          doi = {10.3847/1538-4357/ae42ce},
archivePrefix = {arXiv},
       eprint = {2509.19422},
 primaryClass = {astro-ph.GA},
       adsurl = {https://ui.adsabs.harvard.edu/abs/2026ApJ..1000...90I}
}

@ARTICLE{Wistok2025,
       author = {{Witstok}, Joris and {Jakobsen}, Peter and {Maiolino}, Roberto and {Helton}, Jakob M. and {Johnson}, Benjamin D. and {Robertson}, Brant E. and {Tacchella}, Sandro and {Cameron}, Alex J. and {Smit}, Renske and {Bunker}, Andrew J. and {Saxena}, Aayush and {Sun}, Fengwu and {Alberts}, Stacey and {Arribas}, Santiago and {Baker}, William M. and {Bhatawdekar}, Rachana and {Boyett}, Kristan and {Cargile}, Phillip A. and {Carniani}, Stefano and {Charlot}, St{\'e}phane and {Chevallard}, Jacopo and {Curti}, Mirko and {Curtis-Lake}, Emma and {D'Eugenio}, Francesco and {Eisenstein}, Daniel J. and {Hainline}, Kevin N. and {Jones}, Gareth C. and {Kumari}, Nimisha and {Maseda}, Michael V. and {P{\'e}rez-Gonz{\'a}lez}, Pablo G. and {Rinaldi}, Pierluigi and {Scholtz}, Jan and {{\"U}bler}, Hannah and {Williams}, Christina C. and {Willmer}, Christopher N.~A. and {Willott}, Chris and {Zhu}, Yongda},
        title = "{Witnessing the onset of reionization through Lyman-{\ensuremath{\alpha}} emission at redshift 13}",
      journal = {\nat},
         year = 2025,
        month = mar,
       volume = {639},
       number = {8056},
        pages = {897-901},
          doi = {10.1038/s41586-025-08779-5},
archivePrefix = {arXiv},
       eprint = {2408.16608},
 primaryClass = {astro-ph.GA},
       adsurl = {https://ui.adsabs.harvard.edu/abs/2025Natur.639..897W}
}

@ARTICLE{Barro2025,
       author = {{Barro}, Guillermo and {P{\'e}rez-Gonz{\'a}lez}, Pablo G. and {Kocevski}, Dale and {Trump}, Jonathan R. and {Dickinson}, Mark and {Arrabal Haro}, Pablo and {Brooks}, Madisyn and {Donnan}, Callum T. and {Dunlop}, James S. and {Finkelstein}, Steven L. and {Franco}, Maximilien and {Gandolfi}, Giovanni and {Giavalisco}, Mauro and {Grogin}, Norman A. and {Hirschmann}, Michaela and {Kartaltepe}, Jeyhan S. and {Koekemoer}, Anton M. and {Larson}, Rebecca L. and {Leung}, Gene C.~K. and {Lucas}, Ray A. and {McGrath}, Elizabeth J. and {Papovich}, Casey and {P{\'e}rez-D{\'\i}az}, Borja and {Somerville}, Rachel S. and {Taylor}, Elizabeth and {Taylor}, Anthony J. and {Tripodi}, Roberta and {Yung}, L.~Y. Aaron and {Wang}, Xin},
        title = "{From ``The Cliff'' to ``Virgil'': Mapping the Spectral Diversity of Little Red Dots with JWST/NIRSpec}",
      journal = {\apj},
         year = 2026,
        month = may,
       volume = {1003},
       number = {1},
          eid = {96},
        pages = {96},
          doi = {10.3847/1538-4357/ae5d2c},
archivePrefix = {arXiv},
       eprint = {2512.15853},
 primaryClass = {astro-ph.GA},
       adsurl = {https://ui.adsabs.harvard.edu/abs/2026ApJ..1003...96B}
}

\allauthors
\end{document}